\documentclass[12pt]{article}
\usepackage{epsfig}
\usepackage{amsmath}
\usepackage{hhline}
\usepackage{amssymb}
\usepackage{times}

\newlength{\dinwidth}
\newlength{\dinmargin}
\def\Journal#1#2#3#4{{#1} {\bf #2} (#3) #4}

\def\ar#1#2#3   {\Journal{\em Ann. Rev. Nucl. Part. Sci.}{\bf#1}{#2}{#3}}
\def\err#1#2#3  {\Journal{\em Erratum}{\bf#1}{#2}{#3}}
\def\ib#1#2#3   {\Journal{\em ibid.}{\bf#1}{#2}{#3}}
\def\ijmp#1#2#3 {\Journal{\em Int. J. Mod. Phys.}{\bf#1}{#2}{#3}}
\def\jetp#1#2#3 {\Journal{\em JETP Lett.}{\bf#1}{#2}{#3}}
\def\mpl#1#2#3  {\Journal{\em Mod. Phys. Lett.}{\bf#1}{#2}{#3}}
\def\nim#1#2#3  {\Journal{\em Nucl. Instrum. Meth.}{\bf#1}{#2}{#3}}
\def\nc#1#2#3   {\Journal{\em Nuovo Cim.}{\bf#1}{#2}{#3}}
\def\np#1#2#3   {\Journal{\em Nucl. Phys.}{\bf#1}{#2}{#3}}
\def\pl#1#2#3   {\Journal{\em Phys. Lett.}{\bf#1}{#2}{#3}}
\def\prep#1#2#3 {\Journal{\em Phys. Rep.}{\bf#1}{#2}{#3}}
\def\prev#1#2#3 {\Journal{\em Phys. Rev.}{\bf#1}{#2}{#3}}
\def\prl#1#2#3  {\Journal{\em Phys. Rev. Lett.}{\bf#1}{#2}{#3}}
\def\ptp#1#2#3  {\Journal{\em Prog. Th. Phys.}{\bf#1}{#2}{#3}}
\def\rmp#1#2#3  {\Journal{\em Rev. Mod. Phys.}{\bf#1}{#2}{#3}}
\def\rpp#1#2#3  {\Journal{\em Rep. Prog. Phys.}{\bf#1}{#2}{#3}}
\def\sjnp#1#2#3 {\Journal{\em Sov. J. Nucl. Phys.}{\bf#1}{#2}{#3}}
\def\spj#1#2#3  {\Journal{\em Sov. Phys. JEPT}{\bf#1}{#2}{#3}}
\def\zp#1#2#3   {\Journal{\em Z. Phys.}{\bf#1}{#2}{#3}}
\def\ejc#1#2#3  {\Journal{\em Eur. Phys. J.}{\bf#1}{#2}{#3}}
\def\jetp2#1#2#3 {\Journal{\em J. Exp. Theor. Phys.}{\bf#1}{#2}{#3}}
\def\cpc#1#2#3 {\Journal{\em Comput. Phys. Commun.}{\bf#1}{#2}{#3}}

\def\gsim{\ \,\lower.25ex\hbox{$\scriptstyle\sim$}\kern-1.30ex%
\raise 0.55ex\hbox{$\scriptstyle >$}\ \,}
\def\lsim{\ \,\lower.25ex\hbox{$\scriptstyle\sim$}\kern-1.30ex%
\raise 0.55ex\hbox{$\scriptstyle <$}\ \,}
\newcommand{\rb}[1]{\raisebox{1.5ex}[-1.5ex]{#1}}

\begin{document}

\begin{titlepage}

\begin{flushleft}
 DESY-99-026 \hfill ISSN 0418-9833\\
 March 1999
\end{flushleft}

%
%
%
%
 
\vspace*{2cm}

\begin{center}
\begin{Large}

{\bf Charmonium Production in Deep Inelastic Scattering \\ at HERA}

\vspace{2cm}

H1 Collaboration

\end{Large}
\end{center}

\vspace{2cm}

\begin {abstract}
\noindent
The electroproduction of $J/\psi$ and $\psi(2S)$ mesons is studied in elastic, quasi-elastic
and inclusive reactions for four momentum transfers $2<Q^2<80\mbox{~GeV}^2$
and photon-proton centre of mass energies $25 < W < 180\mbox{~GeV}$. The data
were taken with the H1 detector at the electron proton collider HERA in the years 1995 to 1997. 
The total virtual photon-proton cross section for elastic $J/\psi$ production
is measured as a function of $Q^2$ and $W$. The dependence of the
production rates on the square of the momentum transfer from the proton ($t$) 
is extracted. Decay angular distributions are analysed and the ratio of 
the longitudinal and transverse cross sections is derived.
The ratio of the cross sections for quasi-elastic $\psi(2S)$
and $J/\psi$ meson production is measured as a function of $Q^2$. 
The results are
discussed in terms of theoretical models based upon perturbative QCD. 
Differential cross sections for inclusive and inelastic production of 
$J/\psi$ mesons are determined and predictions within two theoretical
frameworks are compared with the data, the
non-relativistic QCD factorization approach including colour octet and
colour singlet contributions, and the model of Soft Colour Interactions.

\end{abstract}

\vspace{1.5cm}

\begin{center}
Submitted to Eur.~Phys.~J.~C
\end{center}

\end{titlepage}

 C.~Adloff$^{34}$,                
 V.~Andreev$^{25}$,               
 B.~Andrieu$^{28}$,               
 V.~Arkadov$^{35}$,               
 A.~Astvatsatourov$^{35}$,        
 I.~Ayyaz$^{29}$,                 
 A.~Babaev$^{24}$,                
 J.~B\"ahr$^{35}$,                
 P.~Baranov$^{25}$,               
 E.~Barrelet$^{29}$
% ...

% ...
$ ^{24}$ Institute for Theoretical and Experimental Physics,
          Moscow, Russia \\
 $ ^{25}$ Lebedev Physical Institute, Moscow, Russia$^{f,k}$ \\
% ...
$ ^{28}$ LPNHE, \'{E}cole Polytechnique, IN2P3-CNRS, Palaiseau, France \\
 $ ^{29}$ LPNHE, Universit\'{e}s Paris VI and VII, IN2P3-CNRS,
          Paris, France \\
% ...
$ ^{34}$ Fachbereich Physik, Bergische Universit\"at Gesamthochschule
          Wuppertal, Wuppertal, Germany$^a$ \\
 $ ^{35}$ DESY, Institut f\"ur Hochenergiephysik, Zeuthen, Germany$^a$ \\
% ...
$ ^a$ Supported by the Bundesministerium f\"ur Bildung, Wissenschaft,
        Forschung und Technologie, FRG,
        under contract numbers 7AC17P, 7AC47P, 7DO55P, 7HH17I, 7HH27P,
        7HD17P, 7HD27P, 7KI17I, 6MP17I and 7WT87P \\
% ...
$ ^f$ Supported by the Deutsche Forschungsgemeinschaft \\
% ...
$ ^k$ Supported by Russian Foundation for Basic Research 
       grant no. 96-02-00019 

\newpage


\section{Introduction} \label{introduction}

The high energy electron-proton collider HERA has renewed the interest in the study
of light and heavy vector mesons produced  in processes with quasi real and virtual
photon exchange. Several production mechanisms valid in limited kinematic regions
have been discussed in the literature for such processes and  a unified picture is
not available.  The topics of the present paper are studies of elastic and
inelastic production of $J/\psi$ mesons and of quasi-elastic production of
$\psi(2S)$ mesons for four momentum transfers $2 < Q^2 < 80\mbox{~GeV}^2$ and
photon-proton centre of mass energies $25 < W < 180\mbox{~GeV}$.

Leptoproduction of $J/\psi$ mesons has previously been studied in several
fixed target experiments and at HERA in different kinematic 
regions [1--3]. 
In photoproduction, which corresponds to the limit $Q^2 \simeq 0$, and at
low and medium $Q^2$ the production of $J/\psi$ mesons, $e+p\rightarrow e+
J/\psi + X$, is found to be dominated by processes where the  hadronic
system $X$ is either a proton (elastic $J/\psi$ production) or has a
low mass $M_X$. These processes show characteristics of diffraction as
observed in hadron--hadron interactions and in photoproduction of light
vector mesons at lower energies. However, the experiments H1 and ZEUS have
found that at HERA energies the dependence of the elastic $J/\psi$ cross
section on $W$ in the photoproduction limit is steeper than measured in soft
diffractive processes \cite{H1962,Zeus971}.

\begin{figure}[htbp] \centering
\begin{picture}(15.5,8.3)
  \put(-1.0,2.5){\epsfig{file=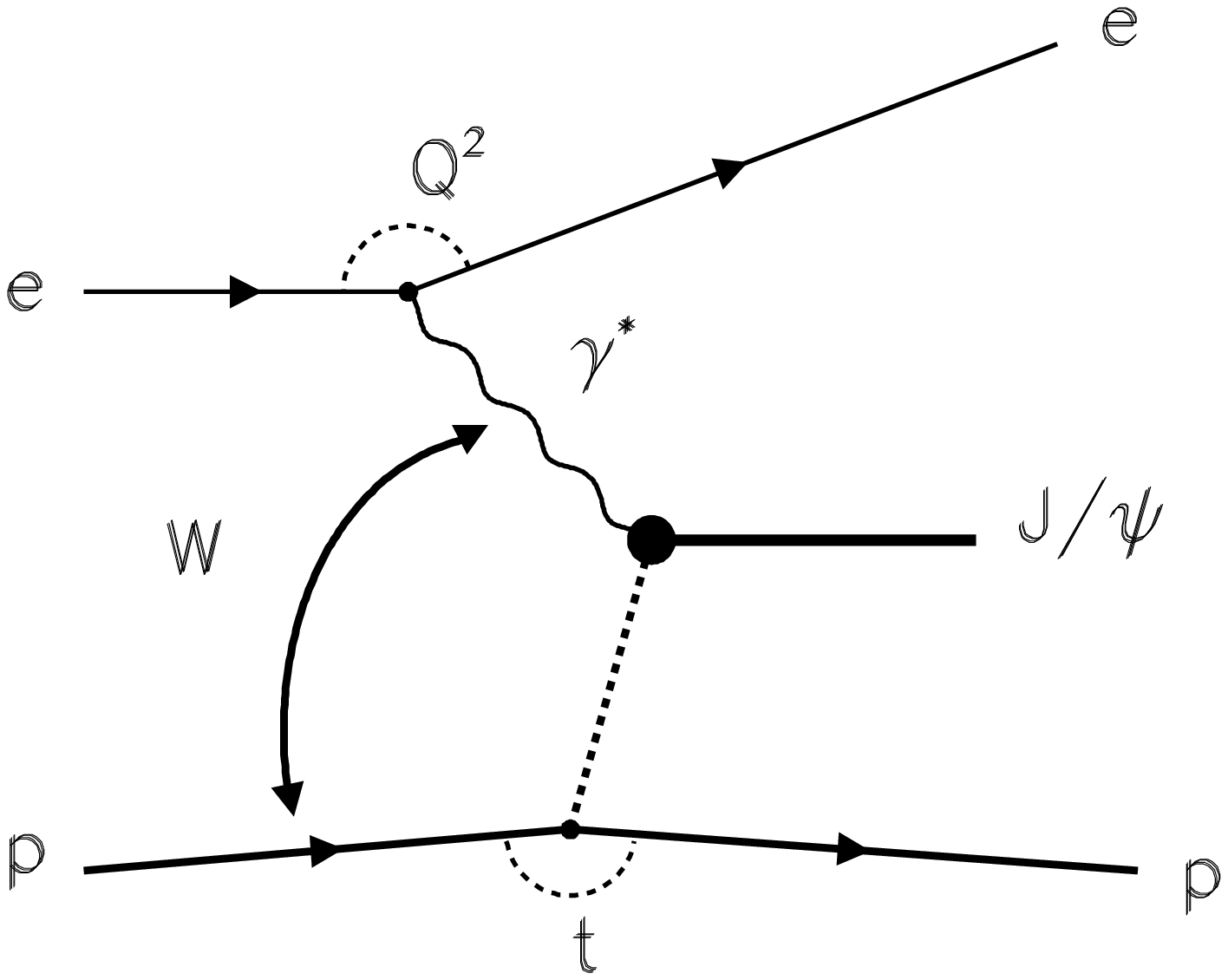,width=6cm}}
  \put(4.5,2.5){\epsfig{file=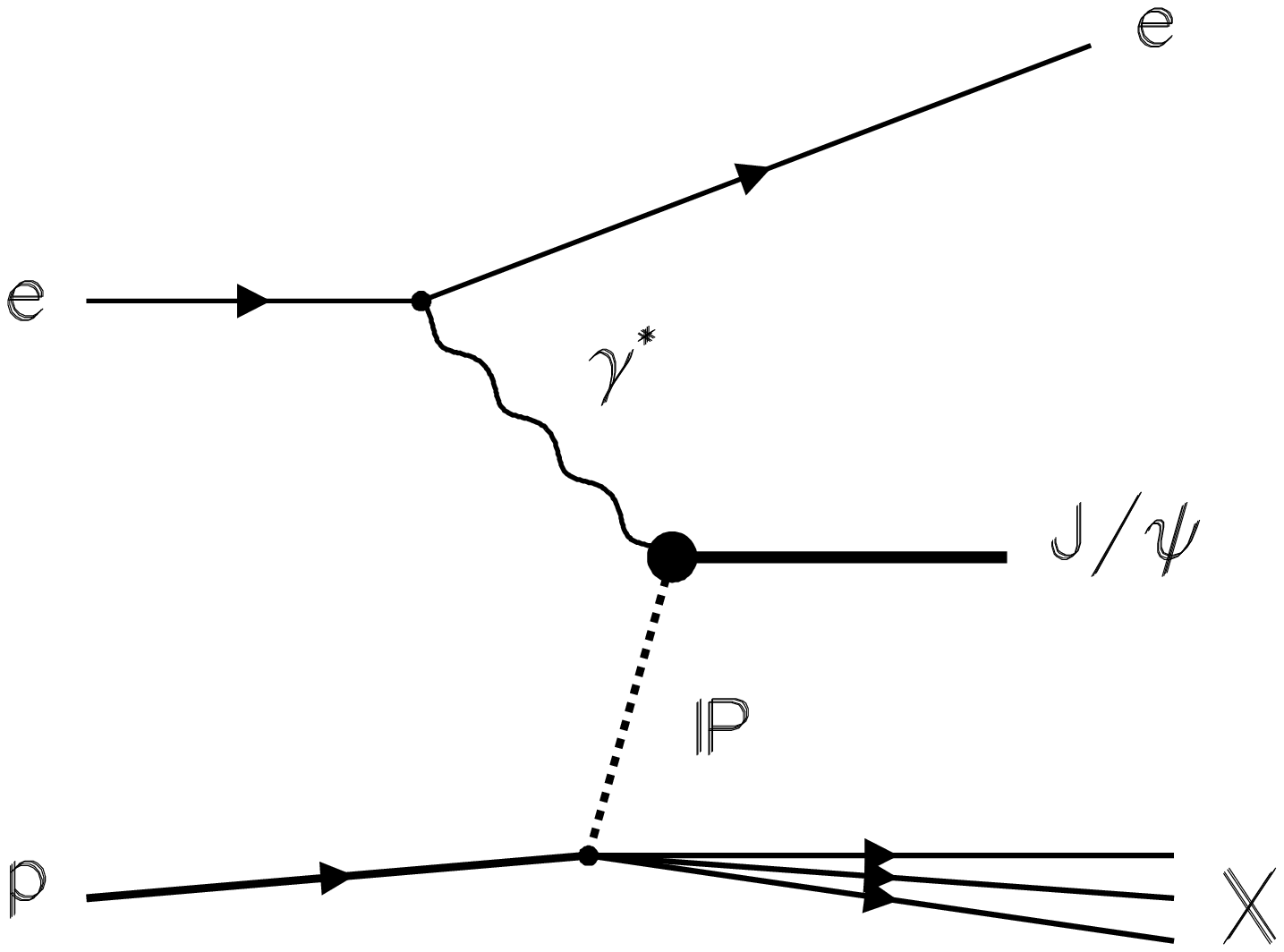,width=6cm}}
  \put(10.0,2.5){\epsfig{file=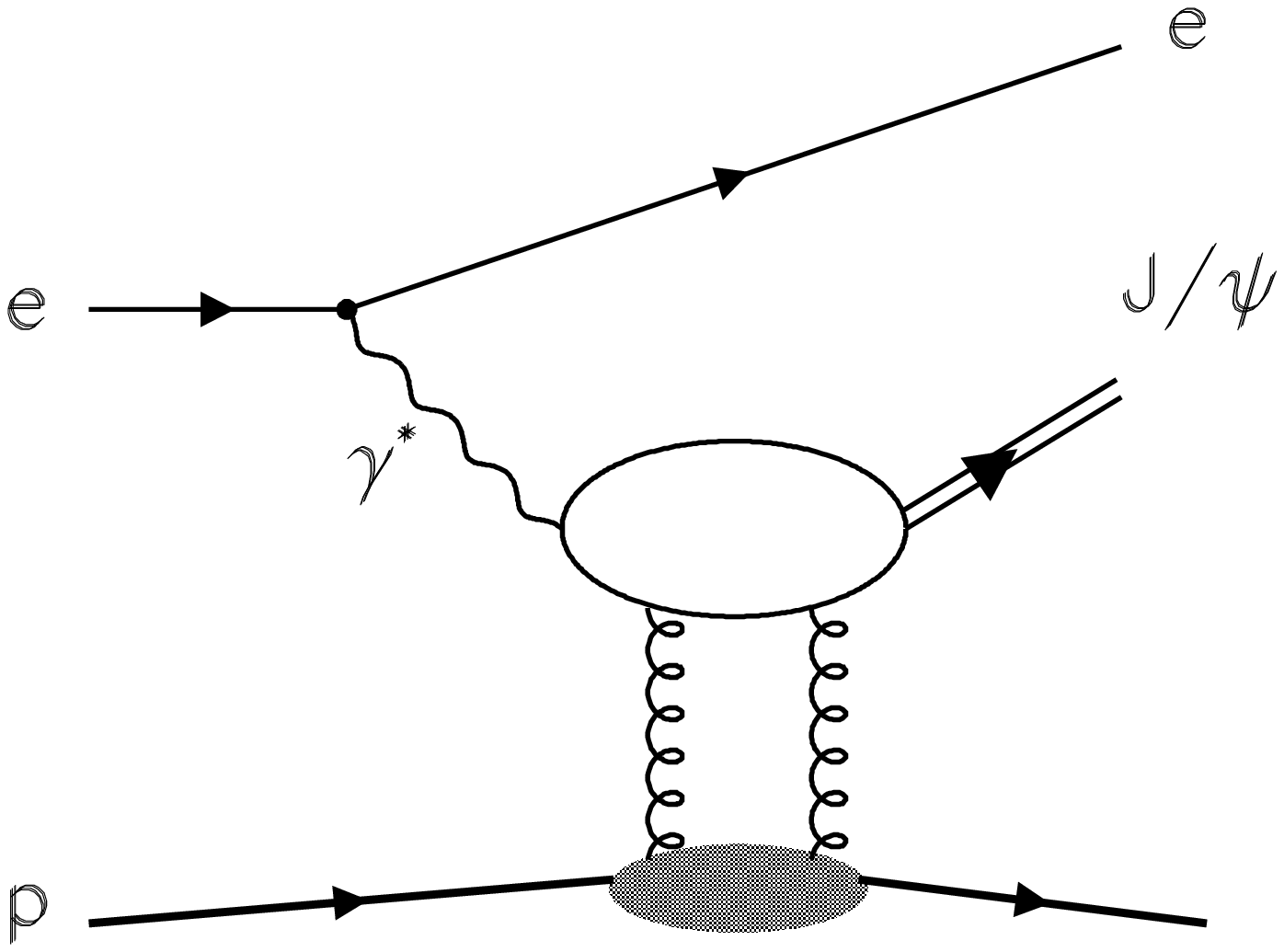,width=6cm}}
  \put(2.0,-1.7){\epsfig{file=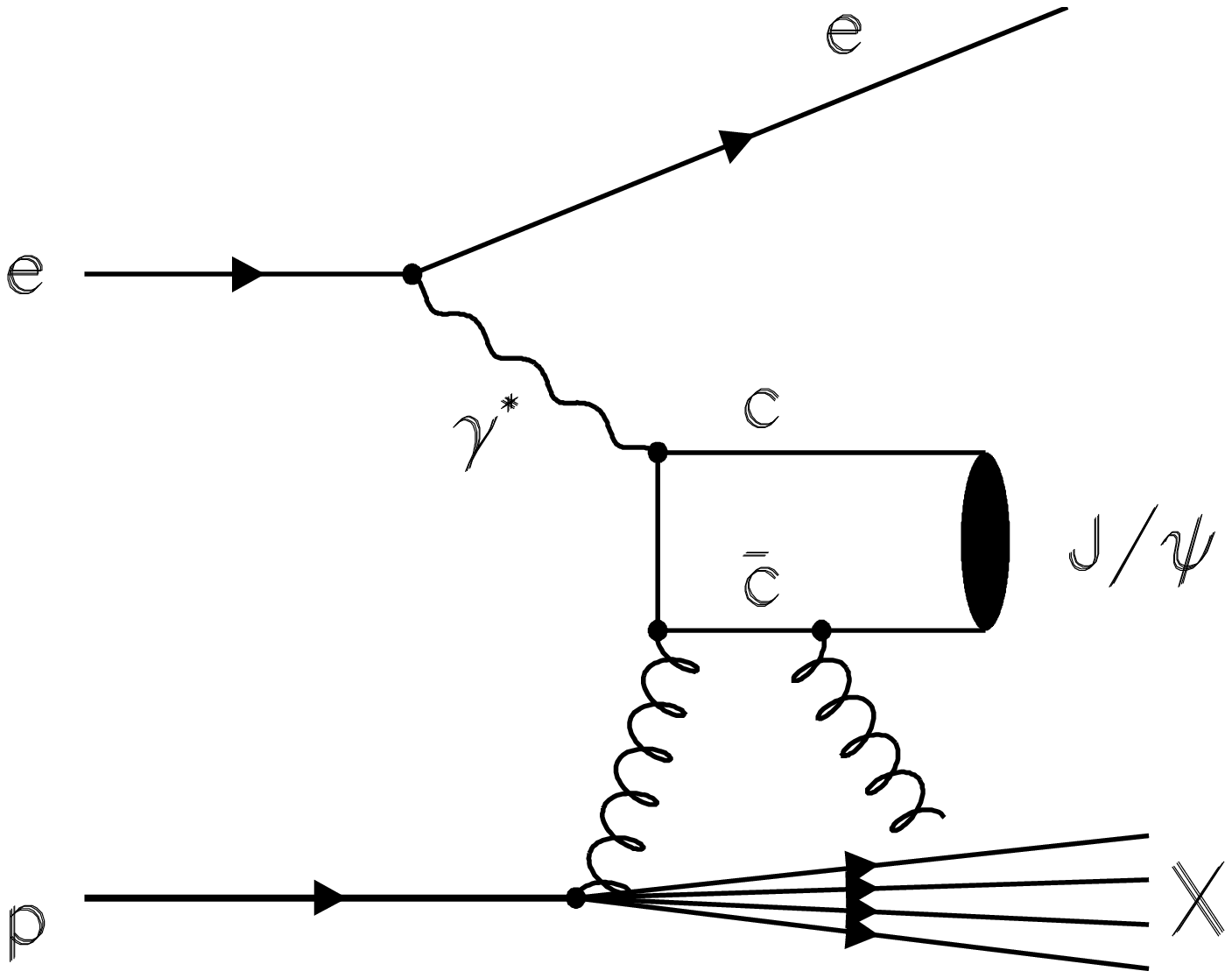,width=6cm}}
  \put(8.0,-1.7){\epsfig{file=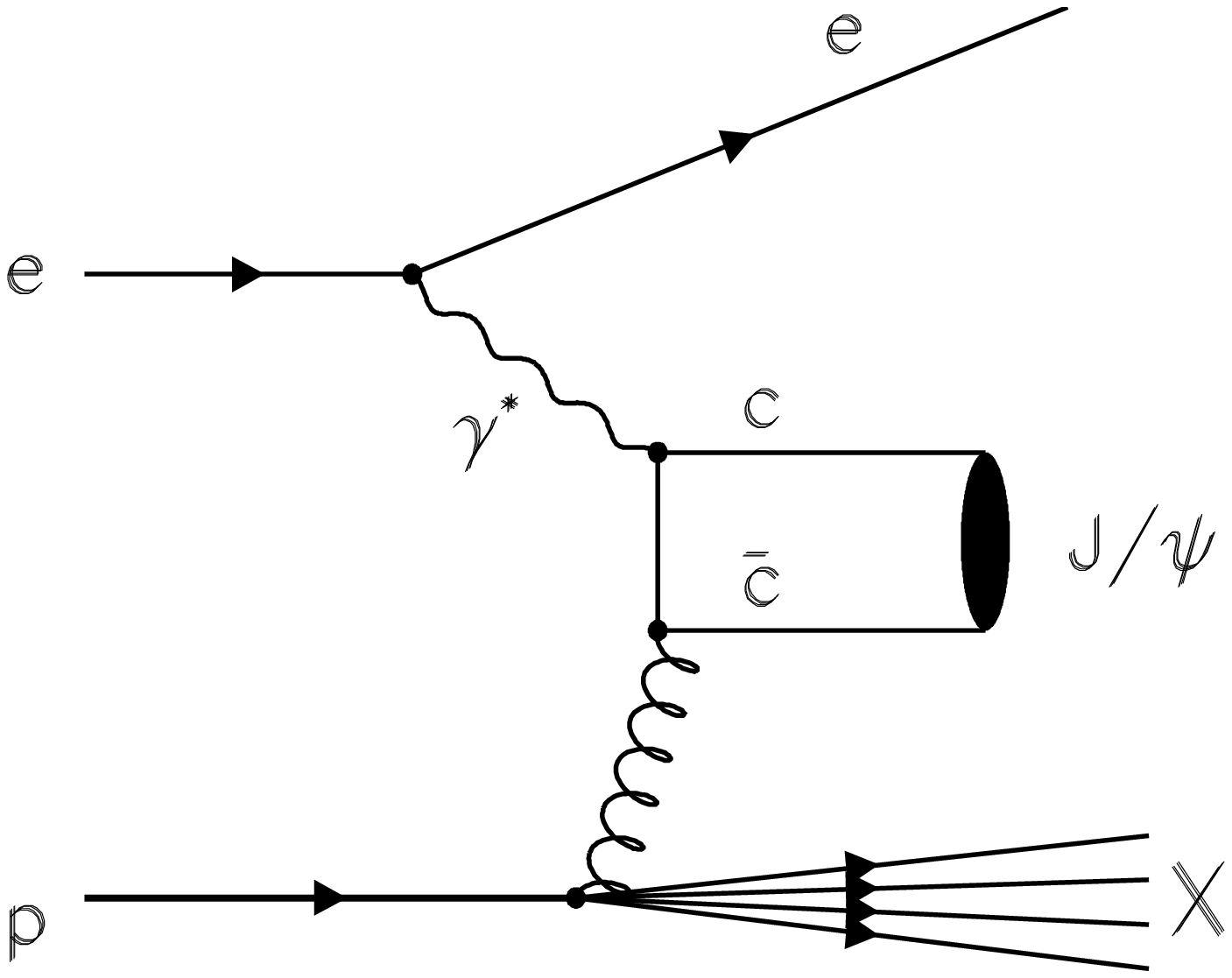,width=6cm}}
  \put(0.0,7.8){\sf a)}
  \put(5.5,7.8){\sf b)}
  \put(11.0,7.8){\sf c)}
  \put(3.0,3.4){\sf d)}
  \put(9.0,3.4){\sf e)}
\end{picture}
\caption{Charmonium production mechanisms: a) elastic and b) 
  proton dissociative production via pomeron exchange;
  c) elastic production via two-gluon exchange; d) leading order diagram in the Colour Singlet
  Model (the $c\bar{c}$ pair is produced in a colour singlet state); e) leading order
  Colour Octet Model (the $c\bar{c}$ pair is produced in a colour octet state).}
\label{diagrams}
\end{figure}

In recent years it has been demonstrated that elastic photo- and electroproduction of
$J/\psi$ mesons (Fig.~\ref{diagrams}a, c) can be calculated within
perturbative QCD (pQCD) [6--9]. 
In these calculations the elastic cross section is related to the square of the
gluon density in the proton and the fast rise of elastic $J/\psi$ production
with $W$ reflects the increase of the gluon density in the proton at low values
of Bjorken $x$ \cite{gluon}. According to these models elastic $J/\psi$ meson
production consequently offers a sensitive way to probe the gluon density.
Further predictions of such models concern, for example, the fraction of
longitudinally polarized $J/\psi$ mesons, the dependences of the slope of the
$t$ distribution ($t$ is the square of the momentum transfer from the proton)
and of the ratio of $\psi(2S)$ to $J/\psi$ meson production on kinematic
variables. The latter ratio is also predicted in an approach based upon colour
dipole phenomenology \cite{Kope}.

Inelastic $J/\psi$ production, which can be described by the formation of
$c\bar{c}$ states via boson gluon fusion (Fig.~\ref{diagrams}d, e), was
previously studied in fixed target experiments \cite{fixtg,nmcemc}
and was analysed in the framework of the Colour Singlet Model \cite{colsing}.
At HERA inelastic $J/\psi$ production has been analysed in photoproduction
\cite{H1962,Zeus972} and was successfully described by predictions of the Colour
Singlet Model in next-to-leading order \cite{kra961}. On the other hand,
measurements of the production of $J/\psi$ mesons in hadronic collisions
\cite{psiteva} have shown that the Colour Singlet Model cannot account for the
observed cross section. A good description of these data can, however, be
achieved using a factorization approach in the framework of non-relativistic
QCD (NRQCD) \cite{nrqcd}, where also colour octet states contribute. These
colour octet processes are also expected to contribute in electroproduction.
First analyses of HERA data showed that the color octet contribution in the
photoproduction regime is less than expected \cite{cogp}. In order to shed
further light on the production process we present a fully inclusive analysis
of $J/\psi$ meson production in the range~$0.2 < z \lsim 1$, where $z$ is the
ratio of the energies of the $J/\psi$ and the exchanged photon in the proton
rest frame. In addition we extract inelastic cross sections in the same $z$
range. The data are compared to calculations of lepton proton scattering in
leading order performed in the NRQCD formalism \cite{Fle972} and to a
phenomenological model incorporating Soft Colour Interactions \cite{sci} in
the Monte Carlo generator AROMA \cite{aroma}.

The paper is organised as follows: after a discussion of different charmonium
production models relevant for the present analysis and a description of the event
selection, the total and differential cross sections for the elastic reaction $e+p
\rightarrow e+J/\psi+p$ are presented with an extended kinematic reach compared to
our previous measurement \cite{H1964} and with statistics increased by an order of
magnitude. We then report on the first measurement of $\psi(2S)$ production in deep
inelastic scattering at HERA and extract the ratio of the $\psi(2S)$ to the
$J/\psi$ meson production cross section as a function of $Q^2$. Finally
differential and total cross sections for inclusive and inelastic $J/\psi$
production are presented.


\section{Models and Phenomenology} \label{models}

The experimental distinction between the various $J/\psi$ production
mechanisms discussed in the literature is not unambiguous and  the following
terminology will be adopted here. The process $$e+p\rightarrow e+ J/\psi +
X$$ will be called ``elastic'' if  $X$ is a proton.  Since the proton is in
general not observed we use the term ``quasi-elastic'' for events in which
only the  tracks of the  $J/\psi$ decay leptons  are present in the main
detector. This data sample comprises in addition  to elastic events those in
which the proton is diffractively excited into a  system $X$ dominantly of
low mass, which subsequently dissociates  (Fig.~\ref{diagrams}b). The decay
or fragmentation products of this low mass system in general escape detection
in the main detector. If a high mass system $X$ is produced the emerging
hadrons are usually detected and the process is called ``inelastic''. The
term ``inclusive'' is used if only the presence of a $J/\psi$ is demanded
irrespective of the production and detection of other particles.

\paragraph{Elastic Charmonium Production}
Elastic photoproduction of light vector mesons, $\rho$, $\omega$ and $\phi$,
is characterized by a weak dependence of the cross section on the photon-proton
centre of mass energy $W$ and by a diffractive peak, i.e.~small
scattering angle of the vector meson with respect to the incident photon
direction. This behaviour is well described by vector meson dominance and
Regge theory in terms of soft pomeron exchange (Fig.~\ref{diagrams}a).
However, the cross section for elastic production of $J/\psi$ mesons by
quasi-real photons ($Q^2\simeq 0$) at HERA is observed to rise steeply with
$W$. Parameterizing the dependence as $W^{\delta}$, $\delta$ is measured to
be of order $1$ for $J/\psi$ mesons \cite{H1962,Zeus971}, while light vector
mesons show an energy dependence compatible with expectations from pomeron
exchange in soft hadronic processes corresponding to $\delta \simeq 0.22 -
0.32$ (see \cite{DL_hadronxsection}).

Modifications of the simple soft pomeron exchange model were subsequently
proposed to describe  the HERA data \cite{pommod}. Alternatively an approach
in the framework of pQCD \cite{Ryskin,Brodsky,F_K_S,F_K_S97} was pursued. In
the models based on pQCD the interaction between the proton and the
$c\bar{c}$ pair is mediated by a system of two gluons (Fig.~\ref{diagrams}c)
or a gluon ladder and the fast increase of the cross section is related to
the rise of the gluon density in the proton at small values of Bjorken $x$.
Since the gluon density enters the cross section quadratically the
sensitivity is large. In these calculations the scattering amplitude is
obtained from the convolution of three contributions which are characterized
by different time scales:  the fluctuation of the (virtual) photon into a
$c\bar{c}$ pair,  the scattering of this hadronic system on the proton,  and
the formation of the final state vector meson.

In contrast to photoproduction where the charm quark mass offers the only hard
scale (at low values of $|t|$),  electroproduction of $J/\psi$ mesons has an
additional scale, $Q^2$, and at high $Q^2$ the predictions of perturbation
theory are expected to  become more reliable. Electroproduction of heavy vector
mesons within pQCD was recently studied in great detail by Frankfurt et
al.~\cite{F_K_S97}. Important corrections  were found concerning, for example,
the choice of the scale at which the gluon density is probed, concerning the
gluon distribution and the wave function for the vector meson and the
importance of corrections due to  Fermi motion of the quarks within the vector
meson.

\paragraph{Inelastic Charmonium Production}\label{secinelmod}
In the  Colour Singlet Model photo- or electroproduction of $J/\psi$ mesons
is assumed to proceed  via boson gluon fusion into  $c\bar{c}$ pairs which
emerge in a colour singlet  state due to the emission of an additional hard
gluon (Fig.~\ref{diagrams}d). The failure of the Colour Singlet Model in
describing  hadroproduction of quarkonia  at large transverse momenta led to
new approaches for the description of $J/\psi$ production which include
contributions from $c\bar{c}$ pairs  in colour octet states. The approach by
Bodwin, Braaten and Lepage (BBL) \cite{nrqcd} based on a factorization
approach in NRQCD was suggested to describe  the large hadroproduction rates
for $J/\psi$ and $\psi(2S)$ production at large $p_t$ measured at the Tevatron
\cite{Bra95}.

When applied to $J/\psi$ electroproduction~\cite{Fle972}, 
the cross section in the BBL formalism  can be expressed as:
\begin{equation}
 \sigma (e+p\rightarrow e+ J/\psi + X) = \sum_n c_n(e+p\rightarrow e+c\bar{c}[n] + X)
             \langle {\cal O}^{J/\psi}_n \rangle , \label{bblxsec}
\end{equation}
where $c\bar{c}[n]$ denotes an intermediate $c\bar{c}$ pair in a definite
colour, spin and angular momentum state $n$. For each $n$, the cross section
factorizes into a short distance part $c_n$ calculable in a perturbative QCD
expansion in the strong coupling parameter $\alpha_s$ and a long distance
matrix  element $\langle {\cal O}^{J/\psi}_n\rangle$ representing the
probability for the $c\bar{c}[n]$ pair to evolve into a colour singlet
$J/\psi$ meson and additional  soft gluons. The long distance matrix elements
are not calculable in perturbation theory and have to be determined
experimentally or  by lattice calculations, but they are thought to be
universal.

The relative importance of the terms in equation (\ref{bblxsec}) is determined 
by NRQCD scaling laws with respect to $v$, the typical relative
velocity of the charm quarks in the $c\bar{c}$ system.
In contrast to the Colour Singlet Model in which all $c_n$ not corresponding 
to colour singlet $c\bar{c}$ states are neglected, the BBL formalism includes 
states where the $c\bar{c}$ system is in a colour octet state (Fig.~\ref{diagrams}e); 
therefore it is often called the ``Colour Octet Model''. However, colour octet
contributions are suppressed by powers of $v^2$. Since $v$ is small, $\langle v^2
\rangle \simeq 0.3$ \cite{nrqcd}, they only become important when the
corresponding short distance coefficients $c_n$ are large. 
In the limit $v\rightarrow 0$ the Colour Singlet Model is restored.

\paragraph{Soft Colour Interactions}
The model of Soft Colour Interactions (SCI) was originally developed as an alternative to
Regge phenomenology and pomeron exchange to describe diffractive scattering at HERA
\cite{edin}. The model was successfully applied to quarkonia production at
the Tevatron \cite{sci}.

For electroproduction the model was implemented in the Monte Carlo program 
AROMA \cite{aroma} which generates $c\bar{c}$ pairs via photon gluon fusion 
according to the leading order matrix elements approximating higher orders by
parton showers. At a scale below the cut-off for pQCD additional interactions take
place: quarks and gluons generated in the hard process interact non-perturbatively
with the partons of the proton remnant; the latter are also allowed to
interact with each other. In these soft colour interactions the momenta of the
partons are not affected, only their colour states may change, which leads to a
modification of the hadronic final state.

In such a model the conversion of a primary $c\bar{c}$ pair
-- which is in a colour octet state -- into an observable colour singlet state
such as $J/\psi$, $\chi_c$, etc.~may occur if the mass corresponds to the
mass of the produced charmonium particle. The
probability for this to happen is not constant as, for example, assumed in the 
Colour Evaporation Model \cite{eva} but depends on the final partonic
state and,  in the Lund string model, on the string configuration.


\section{Detector, Event Selection, Kinematics and Simulations} \label{detector}

The data presented here correspond to an integrated luminosity of $27.3 \pm
0.4\mbox{~pb}^{-1}$. They were collected in the years 1995 to 1997 using the H1
detector which is described in detail in~\cite{detector}. HERA was operated
with $27.5\mbox{~GeV}$ positrons and $820\mbox{~GeV} $ protons.

\subsection{Detector and Event Selection}
$J/\psi$ mesons are detected via the decays $J/\psi \rightarrow \mu^+\mu^-$ and
$J/\psi \rightarrow e^+e^-$ with branching fractions of $6.01\,(6.02) \pm 0.19\%$,
respectively \cite{pdg98}. For $\psi(2S)$ mesons the decay $\psi(2S) \rightarrow
J/\psi \, \pi^+\pi^-$ is used (branching ratio $30.2 \pm 1.9\%$ \cite{pdg98}) with
the subsequent decay of the $J/\psi$ into $\mu^+\mu^-$ or $e^+e^-$. The criteria
for the data selection are summarised in Table \ref{cuts}; further details of the
analysis can be found in \cite{phd}.

In the $Q^2$ range studied here, the scattered positron is identified by its energy
deposition in the backward electromagnetic calorimeter SpaCal \cite{spacal} situated
$152\,\mbox{cm}$ backward from the nominal interaction point of the beams\footnote{H1
uses a right handed coordinate system, the forward ($+z$) direction, with respect to
which polar angles are measured, is defined as that of the incident proton beam, the
backward direction ($-z$) is that of the positron beam.}. The SpaCal covers the polar angles
$155^\circ <\rm \theta < 178^\circ$ and has an energy resolution of 
$\sigma_{E}/E \simeq 7.5\%/\sqrt{E/\mbox{GeV}}\oplus2.5\%$.
A minimal energy deposition of $14\mbox{~GeV}$ is required and cuts are 
applied to the cluster position and cluster shape in order to ensure high
trigger efficiency and a good quality positron measurement. To keep acceptance
corrections small, the $Q^2$ range is limited to $2 < Q^2 < 80\mbox{~GeV}^2$.
A drift chamber (BDC) in front of the SpaCal is used to reconstruct the polar angle 
$\theta_e$ of the scattered positron in combination with the interaction vertex.

The decay leptons of the $J/\psi$ meson are detected in the central 
tracking detector (CTD), consisting
mainly of two coaxial cylindrical drift chambers, which have a length of
$2.2\mbox{~m}$ and outer radii of $0.45\mbox{~m}$ and $0.84\mbox{~m}$. The charged
particle momentum component transverse to the beam direction is measured in these
chambers by the track curvature in the $1.15\mbox{~T}$ magnetic field generated by
the superconducting solenoid, with the field lines directed along the beam axis. Two
polygonal drift chambers with wires perpendicular to the beam direction,
which are located respectively
at the inner radii of the two chambers, are used to improve the
measurement of the particle polar angle. The tracking system is complemented in the
forward direction by a set of drift chambers with 
wires perpendicular to the beam direction which allow particle
detection for polar angles $\theta\gsim 7^\circ$. Multiwire proportional
chambers serve for triggering purposes.

In the present analyses, two oppositely charged tracks with transverse momenta $p_t$
larger than $0.1\mbox{~GeV}$ are required to be reconstructed in the CTD with 
polar angles in the range $20^\circ < \theta < 160^\circ$ where the detection
efficiency is high. For each event, the vertex position in $z$ is determined
using tracks reconstructed in the CTD. To suppress background from 
interactions of the beam
with residual gas in the beam pipe, the vertex must be reconstructed within
$40\mbox{~cm}$ from the nominal interaction point corresponding to $3.5$ times
the width of the vertex distribution.

For the study of elastic $J/\psi$ production to suppress background from inelastic
reactions no track, except a possible track from the scattered positron, is allowed
to be present in addition to the tracks from the two decay leptons.
In the $\psi(2S)$ analysis exactly two tracks, assumed to be pions,
with transverse momenta above $0.12\mbox{~GeV}$ and opposite charge are
required in addition to the two decay leptons (an additional track from the 
scattered positron is allowed). In the study of inclusive $J/\psi$ production
no requirement on the track multiplicity 
is imposed. An inelastic data set is defined by the additional requirement
of an energy deposition of $E_{fwd} > 5\mbox{~GeV}$ in the forward region
of the liquid argon (LAr) calorimeter at polar angles $\theta < 20^\circ$.

\begin{table}[tbp] \centering
\begin{tabular}{l|ll}\hline
  \multicolumn{3}{c}{ } \\
  \multicolumn{3}{c}{\rb{\bf I. Quasi-Elastic \boldmath $J/\psi$ \unboldmath CTD-CTD
      \ \ \ \   $40<W<160\mbox{~\rm GeV}$,\: $2<Q^2<80\mbox{~\rm GeV}^2$}}  \\\hline\hline
                        & \multicolumn{2}{l}{Reconstructed event vertex with $|z_{vtx} - z_{nom}| < 40\mbox{~cm}$ } \\
   {Tracks}             & \multicolumn{2}{l}{Exactly 2 tracks in CTD\footnote{any additional track associated with the scattered~$e^+$ is not considered here.}} \\
                        & \multicolumn{2}{l}{Opposite charges, $20^\circ < \theta < 160^\circ$, $p_t > 0.1\mbox{~GeV}$} \\ \hline
                        & \multicolumn{2}{l}{$\geq 1$ $\mu$ identified in LAr Cal.~(LAr) or Central Muon Detector (CMD) {\em or} \ }\\
  {\rb{Decay leptons}} & \multicolumn{2}{l}{\ \ \ \ 2 $e$ identified in LAr Calorimeter} \\\hline
                        & \rule[0mm]{0mm}{5mm} { Forward untagged:} &
                          \ \ $E_{LAr}^{10^\circ} < 1\mbox{~GeV}$ and  $N_{PRT} = 0$ and $N_{FMD} < 2$ \\
 \rb{Other}            & \rule[-3mm]{0mm}{8mm} { Forward tagged:} &
                          \ \ $E_{LAr}^{10^\circ} > 1\mbox{~GeV}$ or $N_{PRT} > 0$ or  $N_{FMD} \ge 2$ \\\hline\hline
  \multicolumn{3}{c}{ } \\
  \multicolumn{3}{c}{\rb{\bf II. Quasi-Elastic \boldmath $J/\psi$ \unboldmath FMD-CTD (FMD-FMD)
    \ \ \ \             $25<W<40\mbox{~\rm GeV}$,\: $2<Q^2<6\mbox{~\rm GeV}^2$}}  \\\hline\hline
                        &  \multicolumn{2}{l}{1 $\mu$ in FMD and 1 $\mu$ in CTD$+$LAr$/$CMD as in I. {\em or}} \\
   Decay $\mu$          &  \multicolumn{2}{l}{2 $\mu$ in FMD} \\
                        &  \multicolumn{2}{l}{Opposite charges} \\\hline
   Other                &  \multicolumn{2}{l}{No tracks except those 
                   associated with the decay muons $^2$} \\ \hline\hline
  \multicolumn{3}{c}{ } \\
  \multicolumn{3}{c}{\rb{\bf III. Quasi-elastic \boldmath $\psi(2S)$ \unboldmath
       \ \ \ \     $40<W<180\mbox{~\rm GeV}$,\: $1<Q^2<80\mbox{~\rm GeV}^2$}}  \\\hline\hline
                         & \multicolumn{2}{l}{Reconstructed event vertex with $|z_{vtx} - z_{nom}| < 40\mbox{~cm}$ } \\
                         & \multicolumn{2}{l}{Exactly 4 tracks in CTD$^2$}  \\
  {Decay particles}      & \multicolumn{2}{l}{2 $\mu$ identified in LAr or CMD {\em or} 2 $e$ identified in LAr} \\
                         & \multicolumn{2}{l}{Opposite charges of the two leptons and of the two additional tracks (pions)} \\
                         & \multicolumn{2}{l}{$20^\circ < \theta < 160^\circ$, $p_t>800\mbox{~MeV}$ (leptons), $p_t>120\mbox{~MeV}$ ($\pi^+$, $\pi^-$)} \\ \hline\hline
  \multicolumn{3}{c}{ } \\
  \multicolumn{3}{c}{\rb{\bf IV. Inclusive and Inelastic \boldmath $J/\psi$ \unboldmath
     \ \ \ \          $40<W<180\mbox{~\rm GeV}$,\: $2<Q^2<80\mbox{~\rm GeV}^2$}}  \\\hline\hline
  Tracks                 & \multicolumn{2}{l}{Reconstructed event vertex with $|z_{vtx} - z_{nom}| < 40\mbox{~cm}$ } \\ \hline
                         & \multicolumn{2}{l}{2 $\mu$ identified in LAr or CMD {\em or} 2 $e$ identified in LAr} \\
  \rb{Decay leptons}    & \multicolumn{2}{l}{Opposite charges, $20^\circ < \theta < 160^\circ$ and $p_t>800\mbox{~MeV}$}  \\ \hline
  Inelastic selection  &  \multicolumn{2}{l}{$E_{fwd} > 5\mbox{~GeV}$} \\ \hline\hline
  \multicolumn{3}{c}{ } \\
  \multicolumn{3}{c}{\rb{\bf General}}  \\\hline\hline
   Scattered $e^+$      & \multicolumn{2}{l}{\rule[0mm]{0mm}{5mm} $E_e>14\mbox{~GeV}$ identified in SpaCal} \\\hline
   Final state          & \multicolumn{2}{l}{\rule[-3mm]{0mm}{8mm} $\sum (E-p_z) > 45\mbox{~GeV}$ } \\\hline
\end{tabular}
{ \footnotesize $^2$ Any additional track associated with the scattered~$e^+$ is not considered
   here.\hfill}
\caption{Summary of selection criteria for the different data sets.} \label{cuts}
\end{table}

The $J/\psi$ decay leptons are identified by the LAr calorimeter surrounding the
tracking detectors and situated inside the solenoid. The LAr calorimeter is
segmented into electromagnetic and  hadronic sections and  covers the polar
angular range $4^\circ < \theta < 154^\circ$ with full azimuthal coverage.
Muons which are identified as minimum ionizing particles in the LAr  calorimeter
can  in addition be identified by track segments reconstructed in the
instrumented iron return yoke (central muon detector CMD, $4^\circ < \theta
< 171^\circ$) and in the forward muon detector  (FMD, $3^\circ < \theta <
17^\circ$). The FMD provides track segments in front of and behind a toroidal
magnet with a field  of $B = 1.5 - 1.75\mbox{~T}$, thus allowing a determination
of the muon  momentum with a precision of about $25\%$.

The lepton selection criteria vary for the different data sets, depending
on the amount of background. For the quasi-elastic event selection, the identification
of {\em one muon} or {\em two electrons} is required. For the $\psi(2S)$ and the
inclusive data sets two identified leptons with transverse
momenta $p_t > 0.8\mbox{~GeV}$ are required.

A dedicated analysis has been carried out to extend the analysis of elastic
$J/\psi$ production  towards small $W$ ($25 < W < 40\mbox{~GeV}$) using the FMD
(``Low $W$ analysis'').  Two data sets are selected. The events of the first set
which are    required to have  one decay muon in the FMD and the other one in
the CTD (FMD-CTD)  are used for a cross section determination. In the second
data set both muons are   measured in the FMD (FMD-FMD) with an additional loose
vertex requirement  to suppress muons originating from the proton beam halo. The
latter set  ($20 < W < 35\mbox{~GeV}$) serves as a control sample for the FMD
efficiency  which was also determined from a larger photoproduction $J/\psi$
sample. Due to statistical limitations, the low $W$ analysis is restricted to
the low $Q^2$  ($2<Q^2<6\mbox{~GeV}^2$) region.

The triggers in all analyses require a total energy deposition in the
SpaCal above a threshold of $6 - 12\mbox{~GeV}$. The value of the threshold
depends on the topology of the energy deposition and on the presence of
additional requirements, such as signals from the central drift chambers
and/or the multiwire proportional chambers.

In order to minimise the effects of QED radiation in the initial state, the difference
between the total energy and the total longitudinal momentum $\sum (E-p_z)$ reconstructed in
the event is required to be larger than $45\mbox{~GeV}$. If no particle, in particular
a radiated photon, has escaped detection in the backward direction, the
value of $\sum (E-p_z)$ is twice the incident positron energy,~i.e.~$55\mbox{~GeV}$.

\paragraph{Forward Region} \label{forward}
After requiring exactly two tracks corresponding to the $J/\psi$ decay leptons
the event sample includes two main contributions: elastic events  and events
with proton dissociation. It is possible to identify most of the proton
dissociation events with the components of the detector in the forward region,
namely the forward part of the LAr calorimeter, $4^\circ \leq \theta \leq
10^\circ$, the forward muon detector FMD  and the proton remnant tagger (PRT, an
array of scintillators $24\mbox{~m}$ downstream of the interaction point,
$0.06^\circ \leq \theta \leq 0.17^\circ$).  When particles from the
diffractively excited system interact with the material in the beam pipe  or
with the collimators, the interaction products can be detected in these forward
detectors. Events are ``tagged'' as candidates for proton dissociation by the
presence of a cluster  with energy $E_{LAr}$ larger than $1\mbox{~GeV}$ at an
angle $\theta < 10 ^\circ$ in the LAr calorimeter, or by at least 2 pairs of
hits in the first three layers of  the FMD ($N_{FMD} \ge 2$),  or by at least
one hit in the proton remnant tagger ($N_{PRT} > 0$). The forward detectors are
sensitive to  $M_X \gsim 1.6\mbox{~GeV}$.

\subsection{Kinematics} \label{kinematics}

The kinematics for charmonium production is described with the standard
variables used for deep inelastic interactions,
namely the square of the $ep$ centre of mass energy, $s = (p+k)^2$,
 $Q^2 = -q^2$ and $W = (p+q)^2$, where $k$, $p$ and
$q$ are the four-momenta of the incident positron and proton and
of the virtual photon. In addition, the scaled energy transfer $y=p\cdot q / p
\cdot k $ (energy fraction transferred from the positron to the hadronic final state
in the proton rest frame) and the Bjorken variable $x = Q^2 / 2p \cdot q$
are used. Neglecting the positron and proton masses the following relations hold:
$Q^2 = x y s $ and $W^2 = ys - Q^2$.

In the case of elastic $J/\psi$ and quasi-elastic $\psi(2S)$ production the kinematic
variables are reconstructed with the ``double angle'' method \cite{Koijman_Workshop},
where $Q^2$ and $y$ are computed using the polar angles $\theta$ and $\gamma$ of the
positron and of the vector meson in the HERA laboratory frame of
reference, which are well measured:
\begin{equation}
  Q^2 = 4E_0^2 \frac{\sin\gamma (1+\cos\theta)}{\sin\gamma + 
                    \sin\theta - \sin(\gamma + \theta)} , \label{eq:qsq}
\end{equation}
\begin{equation}
  y = \frac{\sin\theta (1 - \cos\gamma)}{\sin\gamma + \sin\theta - \sin(\gamma + \theta)} ; \label{eq:y}
\end{equation}
$E_0$ denotes the energy of the incident positron. The momentum components of
the $J/\psi$ and the $\psi(2S)$ mesons are obtained from their measured decay products. 

Since the fractional energy loss of the proton is negligible,
the absolute value of the four momentum transfer 
$t$ is given to a good approximation by the following relation\footnote{The lowest $|t|$
value kinematically allowed, $t_{min} \simeq (Q^2 + m^2_{V})^2 \ m^2_p\ / y^2 s^2$,
is negligibly small.}:
\begin{equation}
  |t| \simeq (\vec{p}_{tp})^{2} = (\vec{p}_{te} + \vec{p}_{tv})^2,    \label{eq:t}
\end{equation}
where $\vec{p}_{tp}$, $\vec{p}_{te}$ and $\vec{p}_{tv}$ are, respectively, the momentum
components transverse to the beam direction of the final state proton, positron\footnote{The
momentum of the scattered positron is here computed from $Q^2$ and $y$, 
which provides better precision than the direct SpaCal energy measurement.} and
vector meson. 
The resolution obtained from the Monte Carlo simulation for the reconstruction
of $W$ ranges from $4$ to $9\%$ depending on $Q^2$. For $Q^2$ it is 
about $2\%$, and for $t$ on average $0.10\mbox{~GeV}^2$.

The variable $\sum (E-p_z)$ is computed as:
\begin{equation}
  \sum (E-p_z) = (E_e + E_v) - (p_{ze} + p_{zv}),    \label{eq:eminpz}
\end{equation}
$E_e$ and $E_v$ being the measured energies of the scattered positron and of the vector meson,
and $p_{ze}$ and $p_{zv}$ their momentum components parallel to the beam direction.

For inclusive $J/\psi$ production, the ``$e\Sigma$'' method \cite{esigma} is used to
reconstruct the event kinematics, which combines the measurement of both the scattered positron
and the full hadronic final state to obtain good resolution over the entire kinematic
region. The variable $Q^2$ is reconstructed from the scattered positron.
For the calculation of $y$ and  the elasticity $z = (p_\psi\cdot p)/(q\cdot p)$, where $p_\psi$
denotes the $J/\psi$ four-momentum, the observed final state is used in addition. Thus
\begin{equation}
  y = \frac{\sum_{had}(E-p_z)}{\sum (E-p_z)} \mbox{~~~and~~~}
  z = \frac{(E-p_z)_{\psi}}{\sum_{had}(E-p_z)}, \label{yzequation}
\end{equation}
where the sums run over all particles observed in the final state, but excluding the
scattered positron in those indicated by ``had''. For the calculation of the sums in
equation (\ref{yzequation}) a combination of tracks reconstructed in the CTD and cells
in the LAr and SpaCal calorimeters is used. The resolution is good ($2-5\%$) for the
variables $Q^2$, $p_{t,\psi}^2$ and $y^\ast$, the rapidity of the $J/\psi$ in the
$\gamma^{\ast} p$ centre of  mass frame. For $z$ ($W$) the resolution is $2\%$ ($3\%$)
for $z>0.9$ and on average $17\%$ ($9\%$) for $z<0.9$.

\subsection{Monte Carlo Simulations} \label{simulation}

To take account of detector acceptance and efficiencies, smearing effects, losses due
to the selection criteria, and remaining backgrounds, corrections are applied to the
data using Monte Carlo simulations. The H1 detector response is simulated in detail,
and the simulated events are passed through the same reconstruction and analysis
chain as the data.

The correct description of the data by the simulation has been checked
extensively by independent measurements and was adjusted where necessary. In
particular, the trigger efficiencies have been determined using independent
data sets, the efficiency of the central drift chambers has been measured
using cosmic ray muons, and the lepton identification probabilities have been
determined using control samples in which the $J/\psi$ is reconstructed
identifying only one or no lepton. Remaining differences between data and
simulation are used to estimate systematic uncertainties.

Typical efficiencies are:  lepton identification $\sim 80\%$ per lepton,
track reconstruction $\sim 96\%$ per track, and identification of the
scattered positron $99\%$. The total trigger efficiency is determined to
be $97\%$ on average.

The following event generators are used:
\begin{itemize}
  \item The {\bf \mbox{DIFFVM}} program \cite{diffvm} is based on the
    Vector Meson Dominance Model and permits variation of the $Q^2$, $W$ and $t$
    dependences, as well as a variation of the value of $R = \sigma_{L} /
    \sigma_{T}$. In addition to the elastic process, vector meson production with
    proton dissociation is simulated where the dependence of the cross section on
    the mass $M_X$ of the dissociated hadronic state $X$ is parameterized as
    $1/M_X^{2.16}$. High mass states are assumed to decay according
    to the Lund string model \cite{jetset}. In the resonance domain, the mass
    distribution is modelled using measurements from target dissociation on
    deuterium \cite{Goulianos}, and resonance decays are described using their known
    branching ratios.
  \item {\bf EPJPSI} \cite{epjpsi} implements inelastic $J/\psi$ production according
    to the Colour Singlet Model taking into account relativistic corrections and parton showers.
  \item The {\bf LPAIR} generator \cite{lpair} simulates QED electron- and muon-pair
    production, $\gamma \gamma \rightarrow e^+e^-$ and $\gamma \gamma \rightarrow \mu^+\mu^-$,
    where the photons originate from the positron and proton respectively.
    Elastic and inelastic processes are simulated.
\end{itemize}

\paragraph{Radiative Corrections}\label{radcor}
The measured cross sections are given in the QED Born approximation for electron
interactions. The effects of higher order processes -- mainly initial state
radiation -- are estimated using the HECTOR program \cite{hector}.

Radiative corrections for the double angle reconstruction (elastic $J/\psi$ and
quasi-elastic $\psi(2S)$ production) are about $2-3\%$, and are weakly
dependent on $Q^2$ and $W$. A systematic uncertainty of $3\%$ is obtained by
variation of the $Q^2$ and $W$ dependences of the $\gamma^\ast p$ cross 
section within the uncertainties of the measurement. For
the reconstruction of kinematics according to the $e\Sigma$ method (used in the
inclusive $J/\psi$ analysis) the radiative corrections amount to $6-8\%$, 
and again are only weakly
dependent on $Q^2$ and $W$.


\section{Elastic \boldmath $J/\psi$ \unboldmath Production}\label{elastic}

\begin{figure}[b] \centering
\begin{picture}(14.7,6.3)
  \put(0.0,-7.3){\epsfig{file=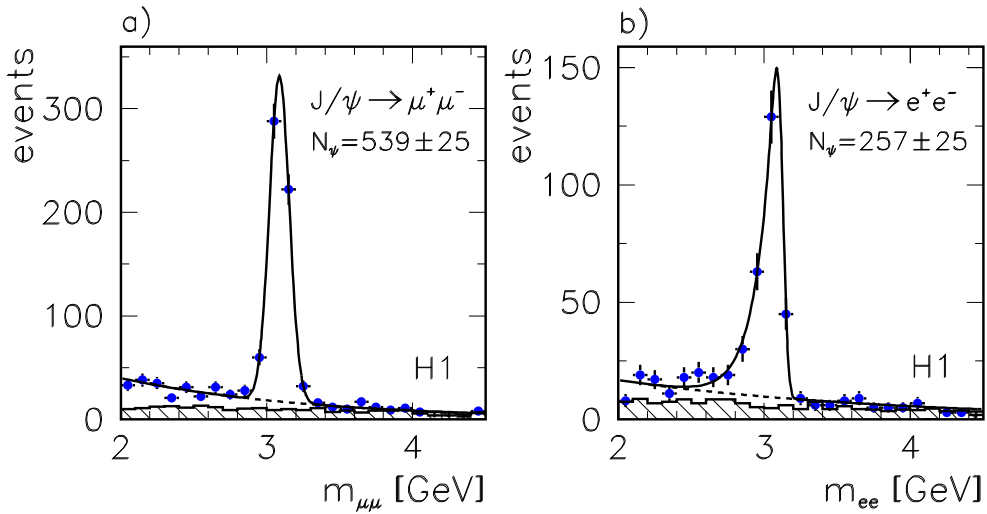,width=14.7cm}}
\end{picture}
\caption{Mass spectra for events of the quasi-elastic $J/\psi$
 selection: a) $\mu^+\mu^-$ pairs, b) $e^+e^-$ pairs. Both
 particles are detected in the central region. The full lines are the results of a fit using a  
 Gaussian distribution for the signal region (convoluted with an exponential tail to account for
 energy loss in the case of di-electron decays) and an exponential distribution for the
 non-resonant background. The mass spectra from two photon processes (LPAIR simulation) are
 shown as hatched histograms. 
 $N_{\psi}$ is the number of $J/\psi$ events obtained from the fit.}
\label{signalsc}
\end{figure}

\begin{figure}
\centering
\begin{picture}(13.6,5.5)
  \put(0.0,-0.9){\epsfig{file=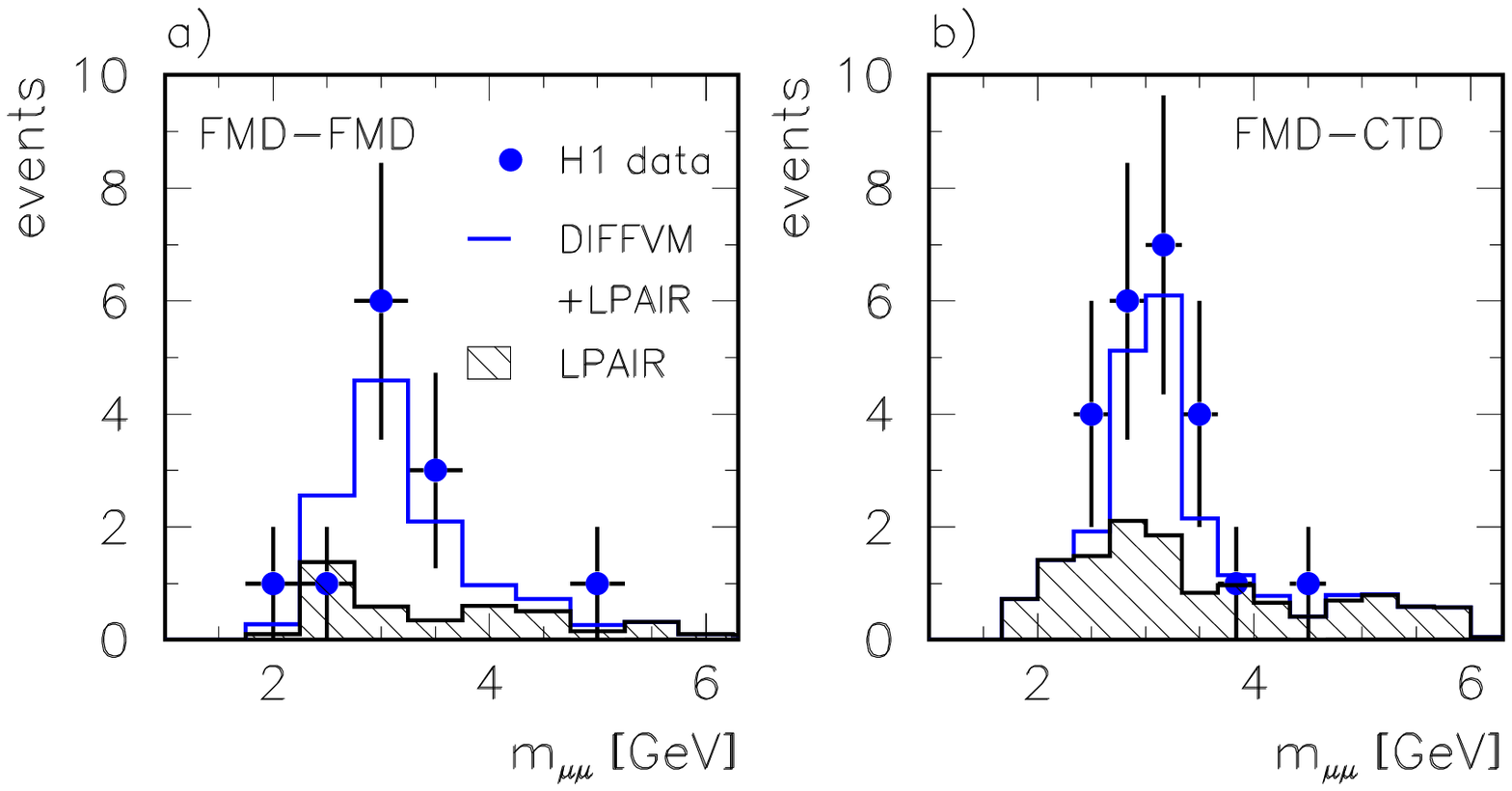,width=13.6cm}}
\end{picture}
\caption{Mass spectra for events of the $J/\psi$ selection in the low $W$ analysis:
  a) $\mu^+\mu^-$ pairs for the FMD-FMD sample and b) for the 
  FMD-CTD sample. The open histogram represents the prediction 
  of a Monte Carlo simulation including diffractively produced $J/\psi$ 
  mesons (\mbox{DIFFVM}) and
  muon pairs from two photon processes (LPAIR) which are also shown 
  separately as hatched histograms.}
\label{signalsf}
 
 \begin{picture}(13.6,12.6)
  \put(0.0,-0.9){\epsfig{file=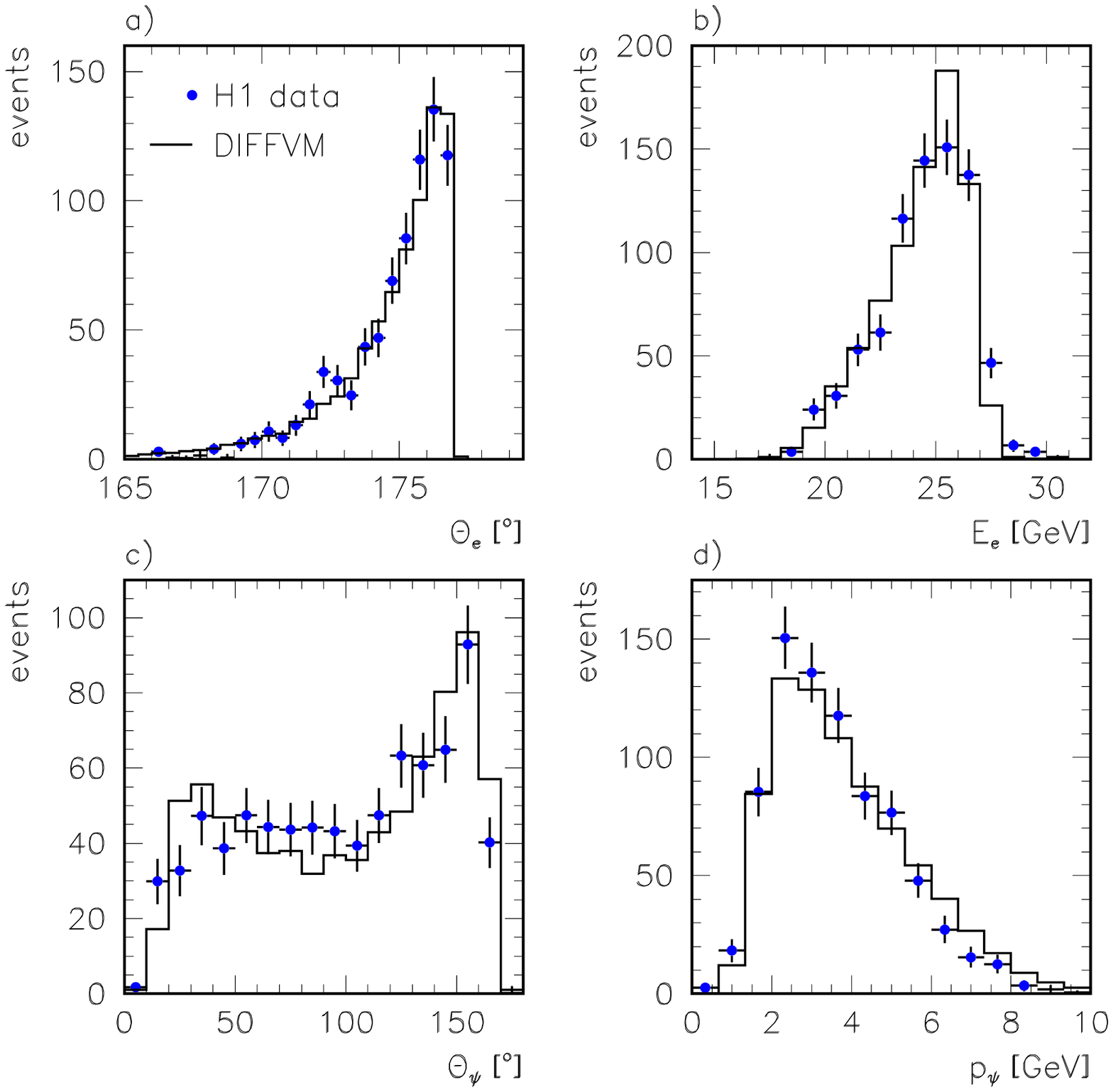,width=13.6cm}}
\end{picture}
\caption{Control distributions for the quasi-elastic $J/\psi$ selection (data set 
  I. in Table \ref{cuts}).
  a) Polar angle and b) energy of the scattered positron, c) polar angle and d)
  momentum of the reconstructed $J/\psi$ candidates. The error bars on the 
  data points are statistical only.
  Results of a diffractive Monte Carlo simulation (\mbox{DIFFVM}) normalised to the 
  data are shown as histograms.}
\label{dataMC}
\end{figure}

The distributions of the invariant mass $m_{ll}$ for the selected events 
in the $J/\psi$ region with two tracks in the central region are presented in
Fig.~\ref{signalsc}a and b, for the $\mu^+\mu^-$ and the $e^+e^-$ decay channels,
respectively (data set CTD-CTD, I. in Table \ref{cuts}) . A clear signal is observed
at $3.090 \pm 0.005\mbox{~GeV}$, compatible with the nominal $J/\psi$ mass of
$3.097\mbox{~GeV}$ \cite{pdg98}. The peak width is compatible with the expectation
obtained from the detector simulation. The mass spectra for the events in the 
low $W$ analysis where one or both muons are reconstructed in the forward muon
detector (FMD) are shown in Fig.~\ref{signalsf}a and b (data sets FMD-FMD and
FMD-CTD, II. in Table \ref{cuts}).

For the events with two tracks in the central detector  the $J/\psi$ signal
region is defined by the condition $| m_{ll} -m_{\psi} | < 250\mbox{~MeV}$,
where $m_{\psi}$ is the nominal $J/\psi$ mass. The non-resonant
background under the $J/\psi$ peak is determined by fitting the sidebands
using an exponential distribution and is found to be $12 \pm 3\%$ on average.
The error includes the uncertainties of the resonance parameterization and of
the background shape where the latter was estimated using a power law as
alternative. The non-resonant background is mainly due to dilepton production
by two photons as simulated by the Monte Carlo generator LPAIR, but there is
also a contribution from hadrons misidentified as leptons. Some distributions
for events in the signal region are shown in  Fig.~\ref{dataMC} as well as
the predictions from a diffractive simulation, \mbox{DIFFVM}. The distributions for
the scattered positron and for the reconstructed  $J/\psi$ meson are
reasonably well described by the simulation. Remaining differences between
data and simulation were checked to have a small impact on the results and are
accounted for by the systematic uncertainty.

\begin{figure}[!b] \centering
\begin{picture}(11.5,10.8)
  \put(0.0,-0.3){\epsfig{file=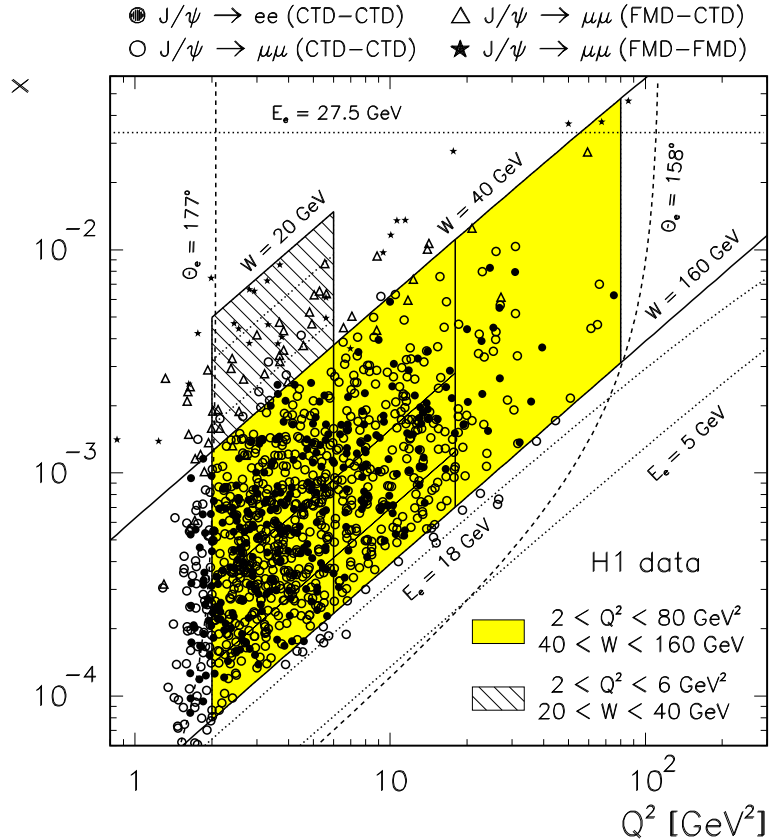,width=11.5cm}}
\end{picture}
\caption{Distribution of quasi-elastic $J/\psi$ candidates in the kinematic 
   ($x, Q^2$) plane (data sets I. and II. in Table \protect\ref{cuts}),
   before applying $Q^2$ and $W$ cuts.
   Lines of constant $W$, energy $E_e$ and polar angle $\theta_e$ of the
   scattered positron are shown and the different analysis regions are indicated.}
\label{kineplane}
\end{figure}

The distribution of the quasi-elastic $J/\psi$ events from the three data
sets (CTD-CTD, FMD-CTD and FMD-FMD) in the kinematic plane $x$ versus
$Q^2$ is displayed in Fig.~\ref{kineplane} before applying $Q^2$ and $W$
cuts.

\subsection{Cross Sections as Functions of \boldmath $W$ and $Q^2$ \unboldmath}

In order to measure the elastic cross section, the quasi-elastic data
sample is divided into two non overlapping classes, {\em forward untagged}
and {\em forward tagged} (see Table \ref{cuts}, I.), which are enriched in
elastic and proton dissociation processes, respectively. The data are
binned in $Q^2$ and $W$ and the non-resonant background is determined and
subtracted for each bin. The ``true'' numbers of elastic and proton
dissociative events $N_{el}$ and $N_{pd}$ are extracted by unfolding them
from the number of events with  and without a tag of the forward
detectors  (see section \ref{forward}). The efficiencies for tagging and
for non-tagging of elastic and proton dissociative events are  determined
by studying the response of the forward detectors and are  incorporated in
the detector simulation. The tagging efficiency for proton dissociative
events with $M_X \gsim 1.6\mbox{~GeV}$ is found to be $92\%$ on average.
Note that in this procedure no assumption is made for the absolute or
relative cross sections of the two processes.

A small correction ($\simeq 3\%$)  due to the presence of spurious hits
in the FMD which are not described by the Monte Carlo simulation is
applied. Further corrections account for the contamination from decays of
the  $\psi(2S)$ meson into a $J/\psi$ and  undetected neutral particles
(based on the results of section \ref{psiprim}) and for initial state
radiation (section \ref{radcor}). Using  the integrated luminosity and
the sum of the branching fractions for the $J/\psi$ meson to decay into
$\mu^+\mu^-$ or $e^+e^-$, an integrated  $ep$ cross section is calculated
for each $Q^2$ and $W$ bin.

In the Born approximation, the electroproduction cross section is related to
the $\gamma^\ast p$ cross section by
\begin{equation}
  \frac{d^2 \sigma (ep \rightarrow e J/\psi\: p)}{dy \ dQ^2} =
  \Gamma \sigma (\gamma ^*p \rightarrow J/\psi\: p) =
  \Gamma \sigma_{T} (\gamma ^*p \rightarrow J/\psi\: p) (1 + \varepsilon R), \label{eq:sigma}
\end{equation}
where $R = \sigma_{L} / \sigma_{T}$, $\sigma_{T}$ and $\sigma_{L}$ are the
transverse and longitudinal $\gamma^\ast p$ cross sections. 
$\Gamma$ is the flux of transverse virtual photons
 \cite{flux} 
and $\varepsilon$ is the flux ratio of longitudinally to transversely polarized
photons, given by
\begin{equation} \Gamma = \frac{\alpha_{em}}{2\pi\: y\: Q^2} \cdot ( 1+ (1-y)^2) ; \label{eq:flux}
    \qquad  \varepsilon = \frac{1 - y}{1-y+y^2/2}.
\end{equation}
Virtual photon-proton cross sections are computed using  equation (\ref{eq:sigma})
after integrating over the $Q^2$ and $W$ bins used in the analysis. The difference
between $\sigma (\gamma ^*p \rightarrow J/\psi p) = \sigma_T + \varepsilon \sigma_L$
and $\sigma_{tot} (\gamma ^*p \rightarrow J/\psi p) = \sigma_T + \sigma_L$ is
negligible here since $\langle\varepsilon\rangle = 0.99$.

For the analysis in which both tracks are detected in the central detector
the systematic uncertainties of the cross sections are estimated to be $17\%$
in total, and are only slightly dependent on the kinematics. They consist of
uncertainties due to detector efficiencies and resolution ($10\%$), uncertainties in
the estimation of background ($11\%$, dominated by proton dissociation and
$\psi(2S)$ decays), radiative corrections and bin centre determination ($4\%$),
the $J/\psi$ decay branching ratio, and luminosity determination  ($4\%$). Part
of the systematic error ($9\%$) affects only the overall normalization. The
uncertainty arising from the proton dissociation background is estimated by
varying the cuts to select proton dissociation events, by changing the $M_X$
dependence assumed in the Monte Carlo simulation, and by changing the model
used for the fragmentation of the system $X$. The uncertainty due to the
subtraction of non-resonant background is determined by varying the assumed
shape of the background and using alternative methods for its determination,
such as sideband subtraction.

\begin{figure}[t] \centering
\begin{picture}(15.0,13.7)
  \put(0.0,-0.9){\epsfig{file=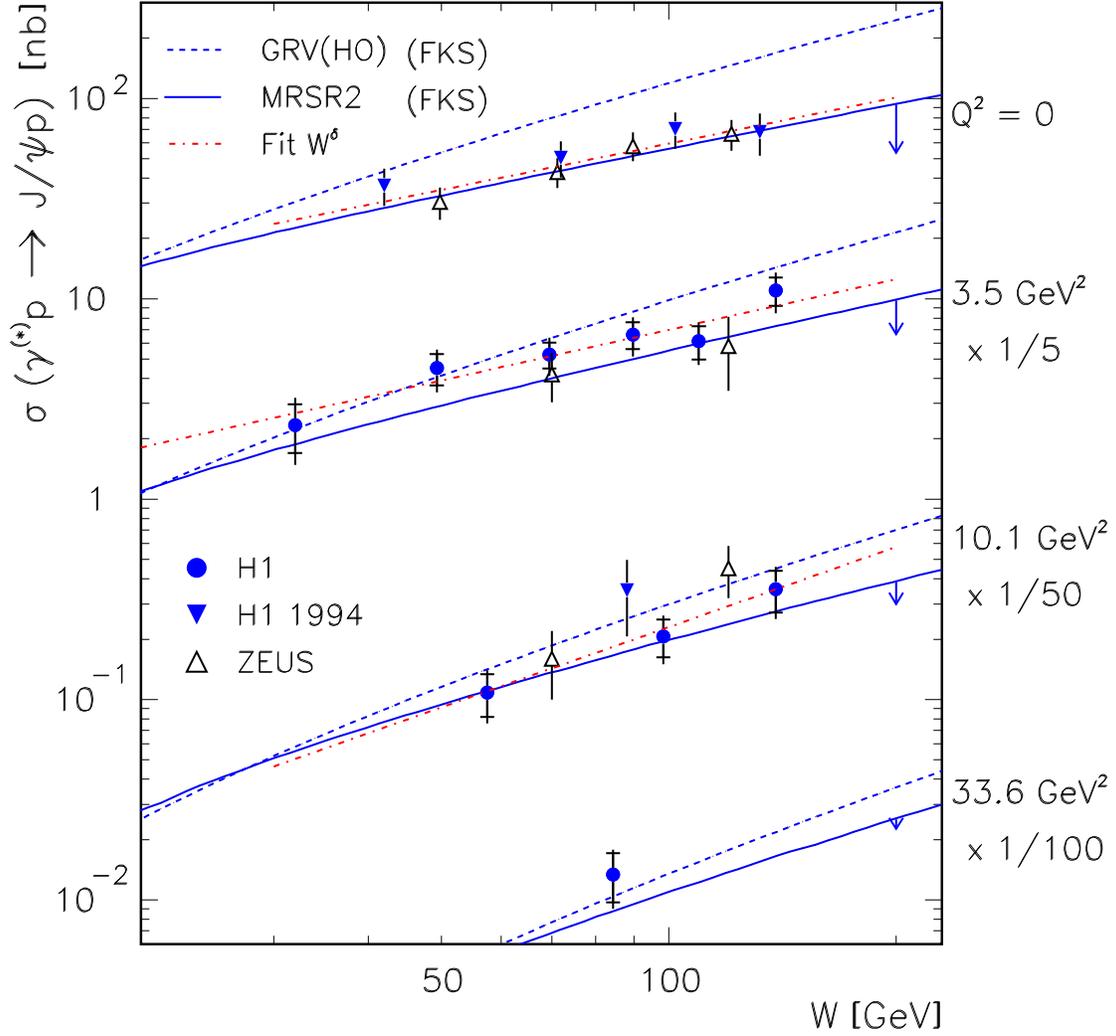,width=15.0cm}}
\end{picture}
\caption{Cross sections for elastic $J/\psi$ production as a function of $W$
  at different values of $Q^2$, measured at HERA in this and other analyses
  \cite{H1964,Zeus981,H1962,Zeus971}. Data for $Q^2 > 0$ have been scaled by factors
  $5$, $50$ and $100$ as indicated. The inner error bars on the points from this
  analysis indicate the statistical errors, while the outer bars show the
  statistical and systematic uncertainties added in quadrature. The dash-dotted
  lines are the results of fits of the form $W^{\delta}$ to the data for each $Q^2$.
  For $Q^2 > 0$ the fits are for H1 data only. The full and dashed lines are results
  of calculations from Frankfurt et al.~\cite{F_K_S97} using different
  parameterizations of the proton gluon densities. The small arrows at
  $W=200\mbox{~GeV}$ indicate the sensitivity of this prediction to a change of the
  charm quark mass from $1.4\mbox{~GeV}$ to $1.5\mbox{~GeV}$.}
\label{sigmaw}
\end{figure}

For the low $W$ analysis (one muon in the FMD) a different procedure to
extract the cross section was adopted due to limited statistics. Since the
contribution of hadrons misidentified as muons is negligible here, the
non-resonant background is subtracted using the LPAIR Monte Carlo
simulation. Alternatively it is estimated from the sidebands of the mass
spectrum. The proton dissociation background is subtracted assuming the
same fraction as determined in the CTD-CTD analysis;  this assumption was
verified by comparing the response of the forward  detectors. The efficiency
of the FMD is determined using a sample of $J/\psi$ photoproduction events
and is cross checked  by the control sample with both muons in the FMD. On
average, the FMD efficiency is found to be $\sim 81\%$. The total
systematic uncertainty of the cross section in the low $W$ analysis is
$25\%$, dominated by the uncertainties in the subtraction of non-resonant
and proton dissociation backgrounds and by the uncertainty of the FMD
efficiency.

The $\gamma^\ast p$ cross sections are  shown in Fig.~\ref{sigmaw} and are given in
Table \ref{summarytable1} as   functions of $W$ in three bins of $Q^2$
($2<Q^2<6\mbox{~GeV}^2$, $6<Q^2<18\mbox{~GeV}^2$ and $18<Q^2<80\mbox{~GeV}^2$). Also
shown are measurements of the ZEUS collaboration\footnote{In the present paper, we
do not use fixed target data for comparison because experimental conditions and
methods  are different and lead to uncertainties in the comparison: for example most
experiments used heavy nuclei as targets and define elastic processes differently.}
at similar values of $Q^2$. The cross sections of the present analysis are quoted at
values of $W$ and $Q^2$  after applying bin centre corrections using the measured
$W$ and $Q^2$ dependences. The $W$ dependence, which in pQCD based models is related
to the $x$ dependence of the gluon density in the proton, is found to be similar to
that  obtained in the photoproduction limit at HERA (also shown in
Fig.~\ref{sigmaw}). When parameterized in the form  $W^\delta$, fits to H1  and ZEUS
photoproduction data yield $\delta=0.77\pm0.18$, while the H1 data for $Q^2 > 0$
yield $\delta=0.84 \pm 0.20$ at $Q^2 = 3.5\mbox{~GeV}^2$ and $\delta=1.3 \pm 0.4$ at
$Q^2 = 10.1\mbox{~GeV}^2$, where the errors include statistical and systematic
uncertainties. The fits are shown in Fig.~\ref{sigmaw}.

In Fig.~\ref{sigmaw} also predictions of the model by  Frankfurt et
al.~\cite{F_K_S97} are included. Gluon densities from GRV(HO) \cite{GRV} and MRSR2
\cite{mrsr2} are used at an effective scale depending on $Q^2$ and on the separation
of the quarks within the $J/\psi$. The prediction using MRSR2 describes the slope of
the data well while the calculation using GRV(HO) is too steep at low values of
$Q^2$. At small $Q^2$, the absolute magnitudes of the predictions are very sensitive
to the input value for the charm quark mass ($m_c = 1.4\mbox{~GeV}$ was chosen
here), as is indicated by the arrows in Fig.~\ref{sigmaw}: For $Q^2 = 0$ and
$W=200\mbox{~GeV}$, for example, a change from $m_c = 1.4\mbox{~GeV}$ to $m_c = 1.5\mbox{~GeV}$
reduces the prediction by more than $40\%$.

\begin{figure}[!t] \centering
\begin{picture}(12.6,11.0)
  \put(0.0,-0.5){\epsfig{file=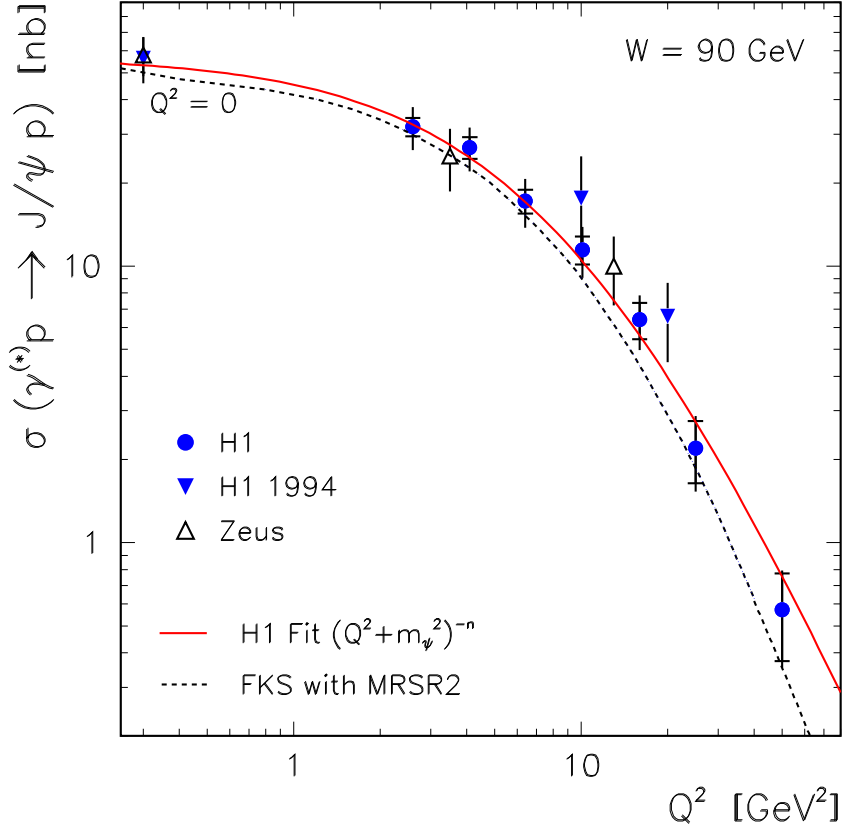,width=12.6cm}}
\end{picture}
\caption{The cross section for elastic $J/\psi$ production at $W = 90\mbox{~GeV}$ as a function
  of $Q^2$. The inner error bars on the points from this analysis indicate the statistical
  errors, while the outer bars show the statistical and systematic uncertainties added in
  quadrature. Also shown are previous measurements in photoproduction (indicated by $Q^2=0$)
  \cite{H1962,Zeus971} and deep inelastic scattering \cite{H1964,Zeus981}. The full line is a
  fit of the form $(Q^2+m_\psi^2)^{-n}$, yielding the result $n=2.38 \pm 0.11$. The dashed
  line is the prediction of Frankfurt et al.~\cite{F_K_S97} using the MRSR2
  \protect\cite{mrsr2} gluon density.}
  \label{sigmaq2}
\end{figure}

The $Q^2$ dependence of the cross section for  $W=90\mbox{~GeV}\,(40 < W <
160\mbox{~GeV})$ is shown in Fig.~\ref{sigmaq2} and given in Table \ref{summarytable2}.
It is well described by a fit $\propto(Q^2+m_\psi^2)^{-n}$ with $n=2.38 \pm 0.11 $. In
order to study a possible change in the observed $Q^2$ dependence, which may be an
indication for the importance of non-perturbative effects, the fits are repeated in two
$Q^2$ regions leading to  $n=2.12\pm0.20$ for $Q^2 < 12\mbox{~GeV}^2$ and $n=2.97\pm0.51$
for $Q^2 > 12\mbox{~GeV}^2$. The errors contain statistical and systematic uncertainties.
In Fig.~\ref{sigmaq2} the model of Frankfurt et al.~with the MRSR2 gluon distribution
which was seen to give a good description of the $W$ dependence (Fig. \ref{sigmaw}) is
also compared to the data. The $Q^2$ dependence is reasonably well described by the
prediction.

\subsection{\boldmath $t$ Distribution and Elastic Slope Parameter \unboldmath}

The elastic slope parameter $b$ is  determined assuming that the $t$
dependence of the elastic $J/\psi$  cross section  can be parameterized by a
single exponential distribution $e^{bt}$. Three contributions are fitted
to the forward untagged  $J/\psi$ sample, corrected for acceptance, losses
and smearing effects. These are:
\begin{itemize}
 \item One exponential distribution $e^{bt}$ with a slope $b$ as free parameter
        describing elastic $J/\psi$ production.
 \item The non-resonant background is described by the sum of two exponential
        distributions contributing in total $12\%$. The $t$-slopes of the non-resonant 
        background are determined using the sidebands of the $J/\psi$ mass distribution.
        The background fraction depends strongly on $t$. For the 
        estimation of the systematic uncertainty, the total amount is varied
        within the range $5-16\%$. 
 \item The proton dissociation background is described by one exponential
       with a slope parameter $1.4\mbox{~GeV}^{-2}$. This is compatible 
       with studies of the forward tagged data set taking into 
       account non-resonant background.
       The total contribution is fixed to $13\%$ and is varied 
       between $5\%$ and $28\%$ to estimate the systematic error, 
       while the slope was varied between $0.8\mbox{~GeV}^{-2}$ and
       $2.0\mbox{~GeV}^{-2}$.
\end{itemize}

\begin{figure}[!t] \centering
\begin{picture}(11.0,9.7)
  \put(0.0,-0.4){\epsfig{file=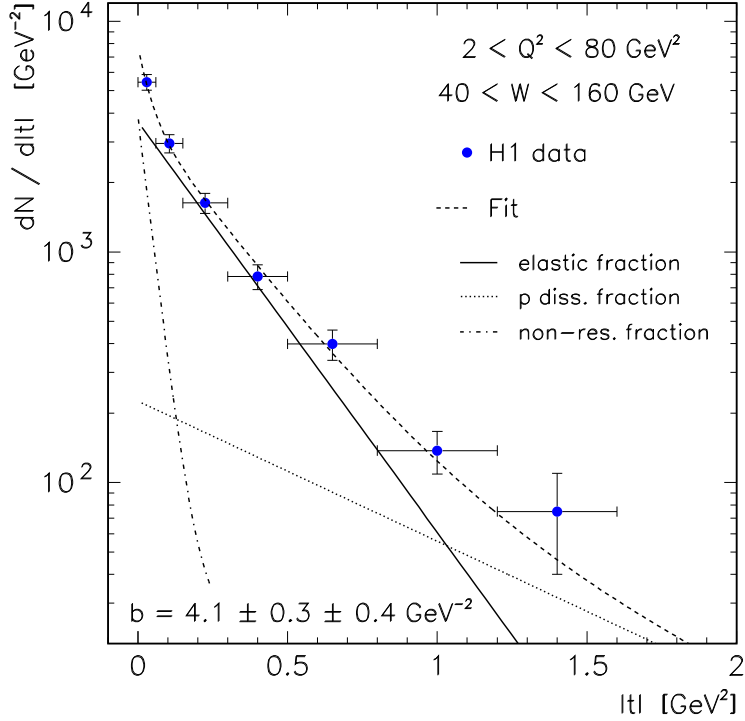,width=11.0cm}}
\end{picture}
\caption{$|t|$ distribution for the forward untagged $J/\psi$ sample,
  corrected for acceptance, losses and smearing effects. The dashed line is
  the result of a fit taking the background contributions into account as described in
  the text. The full line corresponds to the elastic contribution assuming an
  exponential distribution. The contributions from proton dissociation and non-resonant
  background are shown separately. The error bars on the data points are statistical only.}
\label{sigmat}
\end{figure}

No correction for background from $\psi(2S)$ decays is applied  since the total
contribution is small at low $Q^2$.  The result of the fit which is carried out up
to  $|t| = 1.2\,\mbox{~GeV}^2$ is shown in Fig.~\ref{sigmat} ($\chi^2/NDF =
1.6/4$). The elastic slope parameter is
\begin{equation}
 b = 4.1 \pm 0.3\mbox{~(stat.)}\pm 0.4\mbox{~(syst.)}\mbox{~GeV}^{-2}
\end{equation}
for mean values $\langle W\rangle = 96\mbox{~GeV}$ and $\langle Q^2\rangle =
8\mbox{~GeV}^2$. The systematic uncertainty was estimated by varying the fit range
by $\pm 0.4\mbox{~GeV}^2$ and by varying the background contributions and the
corresponding slopes within the ranges given above.

This result for $b$ is compatible with the values obtained by H1 \cite{H1962} and
ZEUS \cite{Zeus971} for elastic $J/\psi$ photoproduction  at similar values of $W$:
$b = 4.4 \pm 0.3\mbox{~GeV}^{-2}$ (H1) and $b = 4.6 \pm 0.6\mbox{~GeV}^{-2}$ (ZEUS),
as well as the ZEUS measurement \cite{Zeus981} of $b = 5.1 \pm 1.3\mbox{~GeV}^{-2}$
for $2<Q^2<40\mbox{~GeV}^2$ and $55<W<125\mbox{~GeV}$. With the present statistics no
significant dependence of the $b$-parameter on $W$ is found (see
Table~\ref{slopetable}); there is however an indication for a decrease of $b$ with
$Q^2$.

\begin{table}[!b] \centering
\begin{tabular}{|c|c|} \hline
  \multicolumn{2}{|c|}{$40 < W < 160\mbox{~GeV}$}                                           \\
  \ $2 < Q^2 < 8\mbox{~GeV}^2$                 & $8 < Q^2 < 80\mbox{~GeV}^2$                 \\\hline
  \ $b = 4.4 \pm 0.4 \pm 0.4\mbox{~GeV}^{-2}$  &  $b = 2.5 \pm 0.6 \pm 0.6\mbox{~GeV}^{-2}$  \\\hline
  \multicolumn{2}{|c|}{$2 < Q^2 < 80\mbox{~GeV}^2$}                                          \\
  \ $40 < W < 100\mbox{~GeV}$                  &    $100 < W < 160\mbox{~GeV}$                \\\hline
  \ $b = 4.0 \pm 0.4 \pm 0.4\mbox{~GeV}^{-2}$  &  $b = 4.1 \pm 0.5 \pm 0.5\mbox{~GeV}^{-2}$   \\\hline
\end{tabular}
\caption{Slope parameters $b$ of the elastic $J/\psi$ meson $t$ distribution 
for different $Q^2$ and $W$ domains.}
\label{slopetable}
\end{table}

\subsection{Decay Angular Distributions for Quasi-Elastic 
            \boldmath $J/\psi$ \unboldmath Production}

In order to investigate the helicity structure of $J/\psi$ meson  production
\cite{Schilling_Wolf}  the angular distributions of the decay leptons in the
helicity frame are used. In this frame, the $J/\psi$ direction in the
$\gamma^\ast p$ centre of mass system  serves as the quantisation  axis. Three
angles are defined: the polar ($\theta^\ast$) and azimuthal ($\varphi$) angles
of the positive decay lepton in the $J/\psi$ rest frame. The third angle  is
the angle $\phi$ between the normals to the $J/\psi$ production plane
(defined by the $J/\psi$ and the scattered proton) and the electron scattering
plane in the $\gamma^\ast p$ centre of mass system.

The one-dimensional distributions in $\cos \theta^\ast$ and
the polarization angle $\Psi=\varphi - \phi$ are extracted. 
If the helicity of the virtual photon is retained by the $J/\psi$ meson
($s$-channel helicity conservation hypothesis, SCHC),
the full angular distribution is a function of $\cos\theta^\ast$ and
$\Psi$ only.

\begin{figure}[!b] \centering
\begin{picture}(13.0,11.5)
  \put(0.0,-0.8){\epsfig{file=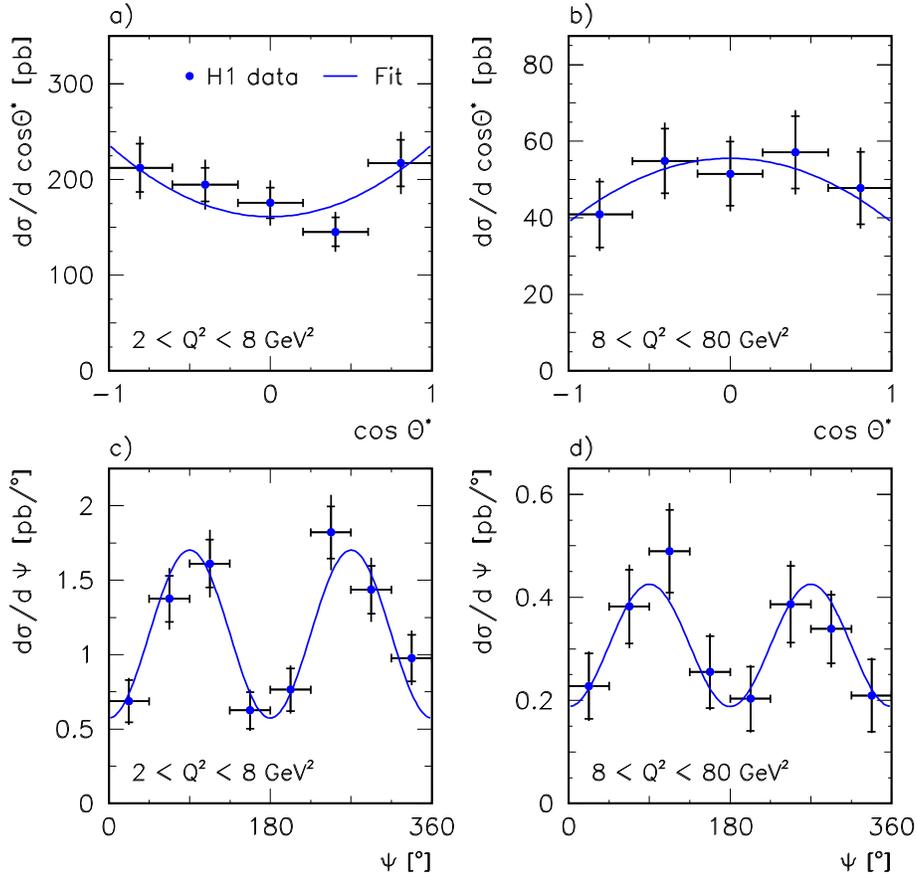,width=13cm}}
\end{picture}
\caption{Angular distributions for the positive $J/\psi$ decay lepton in 
  quasi-elastic production processes $e+p\rightarrow e+ J/\psi + X$ 
  at $40<W<160\mbox{~GeV}$. a) $\cos\theta^*$ for
  $2<Q^2<8\mbox{~GeV}^2$ and b) for $8<Q^2<80\mbox{~GeV}^2$; c) and d) the polarization
  angle $\Psi$ in the same $Q^2$ regions. The inner error bars indicate the statistical
  errors, while the outer bars show the statistical and systematic uncertainties added in
  quadrature. The lines are fits to the data as described in the text.}
\label{sigmahel}
\end{figure}

The acceptance corrected $\cos\theta^\ast$ and $\Psi$ distributions are shown in
Fig.~\ref{sigmahel}. Both the forward tagged and untagged event samples are used,
since the helicity structure is expected to be independent of whether the proton dissociates
or not, a hypothesis supported by the data. The $\cos\theta^\ast$ distribution is
related to the spin density matrix element $r^{04}_{00}$, the probability of
the $J/\psi$ meson to be longitudinally polarized, according to
\begin{equation}
  \frac{{\rm d}\,\sigma}{{\rm d}\,\cos\theta^{\ast}}\propto 1 + r^{04}_{00} + (1 - 3r^{04}_{00}) \cos^2\theta^\ast .
\end{equation}
A fit to the data shown in Figs. \ref{sigmahel} a) and b) yields
\begin{eqnarray}
    r^{04}_{00} & = & 0.15\pm 0.11 \qquad \mbox{for}\qquad \langle Q^2\rangle = 4\mbox{~GeV}^2,\\
    r^{04}_{00} & = & 0.48\pm 0.15 \qquad \mbox{for}\qquad \langle Q^2\rangle = 16\mbox{~GeV}^2 .
\end{eqnarray}

The $\Psi$ distribution is related to the spin density matrix element $r^{1}_{1 -1}$:
\begin{equation}
  \frac{{\rm d}\,\sigma}{{\rm d}\,\Psi} \propto 1 - \varepsilon\, r^{1}_{1 -1} \cos 2 \Psi .
\end{equation}
A fit to the data shown in Figs. \ref{sigmahel} c) and d) yields
\begin{eqnarray}
    r^{1}_{1 -1} & = & 0.50\pm 0.08 \qquad \mbox{for}\qquad \langle Q^2\rangle = 4\mbox{~GeV}^2,\\
    r^{1}_{1 -1} & = & 0.39\pm 0.13 \qquad \mbox{for}\qquad \langle Q^2\rangle = 16\mbox{~GeV}^2 .
\end{eqnarray}

In the case of $s$-channel helicity conservation and natural parity exchange 
(NPE) the matrix elements $r^{04}_{00}$ and $r^{1}_{1 -1}$ are related by 
$  r^{1}_{1 -1} = \frac{1}{2} ( 1 - r^{04}_{00} ).$
Using this relation and the measured values for $r^{04}_{00}$, one obtains 
values for $r^{1}_{1 -1}$ which agree to within one standard deviation
with those obtained from the $\Psi$ angular distributions, thus supporting
the SCHC and NPE hypotheses.

Under the assumption of SCHC the measurement of the $r^{04}_{00}$
matrix element can be used for the determination of $R$, the ratio
of the longitudinal to the transverse cross section:
\begin{equation}
  R = \frac{\sigma_L}{\sigma_T} = \frac{1}{\varepsilon} \ \frac{r^{04}_{00}}{1-r^{04}_{00}}.
\end{equation}
Using this relationship and the $\cos\theta^*$ distribution $R$ is determined
in two $Q^2$ regions:
\begin{eqnarray}
       R &= & 0.18^{+0.18}_{-0.14}\qquad \mbox{for}\qquad \langle Q^2\rangle = 4\mbox{~GeV}^2,\\
       R &= & 0.94^{+0.79}_{-0.43}\qquad \mbox{for}\qquad \langle Q^2\rangle = 16\mbox{~GeV}^2 .
\end{eqnarray}
A measurement of $R = 0.41 ^{+0.45}_{-0.52}$ by the ZEUS experiment \cite{Zeus981} at
$\langle Q^2\rangle = 5.9\mbox{~GeV}^2$ and $\langle W\rangle =97\mbox{~GeV}$
is compatible with these values. Taking into account the photoproduction measurements of
$R = 0.17 \pm 0.14$ \cite{H1962} and $R = -0.01 \pm 0.09$ \cite{Zeus971}, which
are compatible with the expectation $R = 0$ for $Q^2 = 0$, a rise of
$R$ with increasing $Q^2$ is suggested by the data.

The measured values of the $R$ parameter are significantly
smaller for $J/\psi$ than for elastic $\rho$ meson production at HERA \cite{H1964,Zeus981,H1991}
at similar $Q^2$; but they are of the same order 
if compared at the same value of $Q^2/m_V^2$, where $m_V$ is the mass of the $\rho$
or the $J/\psi$.


\section{\boldmath Quasi-elastic $\psi(2S)$ Production \unboldmath}\label{psiprim}

For the selection of $\psi(2S)$ mesons the decay channel $\psi(2S) \rightarrow
J/\psi\,\pi^+\pi^-$, where the $J/\psi$ decays either in two electrons or two
muons, is used. In this case no separation between elastic and proton dissociation is
attempted due to limited statistics. The goal is to derive the ratio of cross sections
for $J/\psi$ and $\psi(2S)$ production as a function of $Q^2$. The lower $Q^2$ cut is
reduced to $1\mbox{~GeV}^2$ since the $Q^2$ dependent acceptance corrections cancel
almost completely in the cross section ratio.

\begin{figure} \centering
\begin{picture}(13.0,7.0)
  \put(0.0,-6.2){\epsfig{file=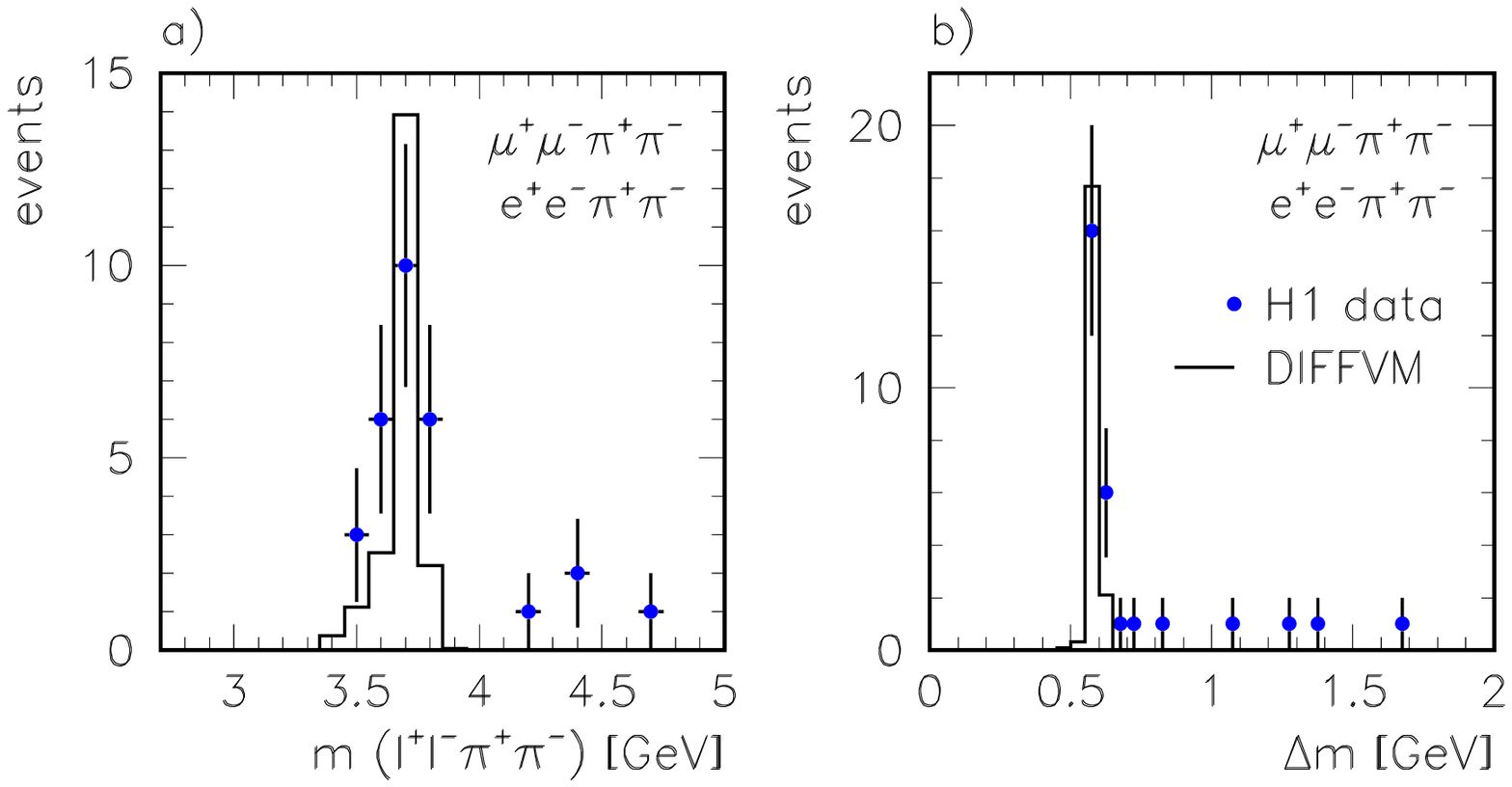,width=13cm}}
\end{picture}
\caption{a) Mass spectrum of the four particles $\ell^+\ell^-\pi^+\pi^-$ 
  and b) $\Delta m = m_{\psi(2S)}-m_{\psi}$
  for the $\psi(2S)$ candidate events, i.e.~for events with $|m_{ll} - m_{\psi}| < 300\mbox{~MeV}$.
  The \mbox{DIFFVM} Monte Carlo simulations for the signals are shown for comparison.}
\label{signals2}

\begin{picture}(10.0,10.0)
  \put(0.0,-0.3){\epsfig{file=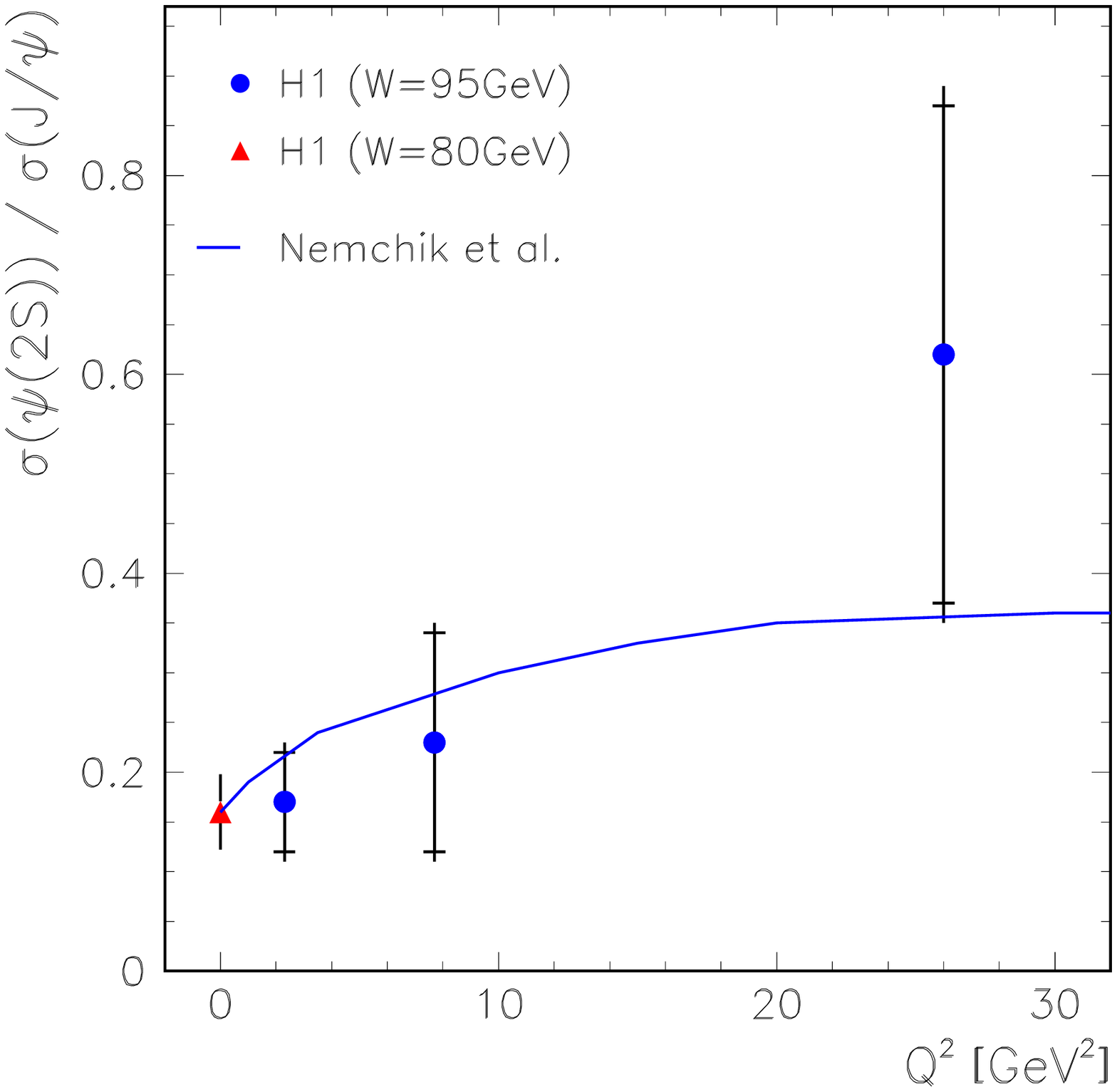,width=10cm}}
\end{picture}
\caption{Ratio of cross sections for the quasi-elastic  production of 
  $\psi(2S)$ and $J/\psi$ mesons 
  as a function of $Q^2$ for this analysis and for the H1 photoproduction measurement
  \protect\cite{H1963}, corrected for the most recent
  branching fraction $BR(\psi(2S) \rightarrow J/\psi \, \pi^+\pi^-) =
  30.2 \pm 1.9\%$ \cite{pdg98}. The inner error bars on the H1 points from this analysis
  indicate the statistical uncertainty, while the outer bars show the statistical and systematic
  uncertainties added in quadrature. The prediction from \cite{Kope} based on
  colour dipole phenomenology is also displayed.}
\label{ratio}
\end{figure}

The signals in the quasi-elastic $\psi(2S)$ selection are displayed in
Fig.~\ref{signals2}. 
For the determination of the $\psi(2S)$ to $J/\psi$ ratio the
non-resonant background is subtracted using the sidebands of the di-lepton mass
spectrum in the case of the $J/\psi$ meson and using the sidebands of the 
$\Delta m = m_{\psi(2S)}-m_{\psi}$ distribution for the $\psi(2S)$ meson. The data are divided
in three $Q^2$
bins: $1<Q^2<5\mbox{~GeV}^2$, $5<Q^2<12\mbox{~GeV}^2$ and $12<Q^2<80\mbox{~GeV}^2$.
The cross section ratio is shown in Fig.~\ref{ratio}. The systematic uncertainty of the ratio 
amounts to $16\%$ in total and is dominated by the contribution from the track
reconstruction efficiency for the low momentum $\pi^+\pi^-$ pair.  

The measurement at low $Q^2$ agrees well with the H1 photoproduction measurement
\cite{H1963}. An indication of a rise with $Q^2$ at the level of two standard
deviations is observed which is also predicted in models by Frankfurt et
al.~\cite{F_K_S} and Nemchik et al.~\cite{Kope}. In ref.~\cite{F_K_S} an
asymptotic value of $\sigma_{\psi(2S)}/\sigma_\psi \approx 0.5$ is expected for
$Q^2 \gg m_\psi^2$.


\section{Inclusive and Inelastic \boldmath $J/\psi$ Production \unboldmath}

Inclusive $J/\psi$ production is studied in the kinematic range $2 < Q^2 <
80\mbox{~GeV}^2$ and $40 < W < 180\mbox{~GeV}$ covering $0.2 < z \lsim 1.0$ for
the muonic decay of the $J/\psi$, while for the decay to electrons $z > 0.5$
is required. The restricted $z$ region for $J/\psi\rightarrow e^+e^-$ is due to
the smaller acceptance for electrons and larger background at low $z$  values.
The elasticity $z$ (defined in section \ref{kinematics}) denotes the ratio of
energies of the $J/\psi$ and of the exchanged photon in the proton rest frame.
Two sets of differential cross sections are determined. First an {\em inclusive}
cross section is derived where in the given kinematic region all $J/\psi$ mesons are
selected irrespective of the production mechanism, thus including inelastic and
elastic contributions. The inclusive cross sections are compared to the
predictions of the Soft Colour Interaction Model \cite{sci}. A second set of
differential cross sections is derived for {\em inelastic} $J/\psi$
production which can be compared to the predictions within the NRQCD factorization
approach \cite{Fle972} containing colour octet contributions.

An ``inelastic'' production process can be defined experimentally in several
ways. In previous photoproduction analyses \cite{H1962,Zeus972} cuts in the
variable $z$, e.g. $z\lsim 0.9$, were used to suppress elastic and proton
dissociative events. In the present analysis a different approach is chosen
because colour octet contributions are, in leading order in $\alpha_s$,
predominantly expected at large $z$. This is because the $c\bar{c}$ pair can
be produced with no other particles in the final state, i.e.~$z\sim1$ (see
Fig.~\ref{diagrams}e). Its non-perturbative evolution into the $J/\psi$ meson
reduces the value of $z$ only slightly, and applying a $z$ cut as done
previously would reduce the expected colour octet contributions together with
the quasi-elastic ones by an unknown amount. In the present analysis a cross
section is determined suppressing contributions of low mass $M_X$ following a
suggestion of \cite{Fle972}. This suppression of low masses is achieved
indirectly by requiring a minimal calorimetric energy in the forward region
of the detector. This requirement selects high masses and suppresses elastic
and proton dissociative events characterised by small  $M_X$ corresponding to
small energy deposits in the forward direction. Colour octet contributions are
expected to have $M_X \gsim 15\mbox{~GeV}$ \cite{Fle972} and are retained.

\subsection{Data Analysis}
\begin{figure} \centering
\begin{picture}(13.6,5.7)
  \put(0.0,-0.9){\epsfig{file=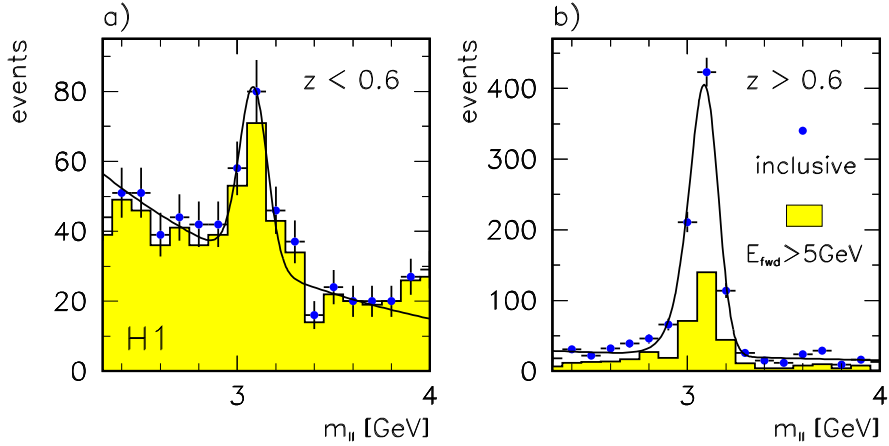,width=13.6cm}}
\end{picture}
\caption{Di-lepton mass spectra for events of the inclusive (points) and inelastic (histogram) $J/\psi$
 selection: a) $0.2 < z < 0.6$, b) $0.6 < z \lsim 1$. The curves are the results of fits of
 Gaussian distributions for the signal (convoluted with an exponential tail to account for
 energy loss in the case of di-electron decays) and an exponential distribution for the
 non-resonant background.}
\label{signalsi}
\end{figure}

The selection criteria as described in Table~\ref{cuts} are used. The
di-lepton mass spectra of the selected events are shown in
Fig.~\ref{signalsi} separately for $0.2 < z < 0.6$ and $0.6 < z \lsim 1$,
both for the inclusive and the inelastic selection ($E_{fwd} >
5\mbox{~GeV}$). Since the non-resonant background increases with decreasing
$z$ the background fraction is determined from the mass spectra in bins of
$z$ by fitting the signal and background as in  section \ref{elastic}. For
the determination of the differential cross sections a correction is applied
according to the $z$ values of the events.

Acceptance and efficiencies are determined using a simulation tailored to describe the
data, which consists of a mix of diffractive events generated by \mbox{DIFFVM} \cite{diffvm}
and of inelastic events generated by EPJPSI \cite{epjpsi}. The diffractive events are
composed of  elastic and proton dissociative contributions in a ratio consistent with
the signals observed in the forward detectors (compare section \ref{elastic}). EPJPSI
generates events according to the Colour Singlet Model. Both models were previously
shown to describe quasi-elastic and inelastic data respectively (see for
example~\cite{H1962}).  Contributions from other processes such as $b$-decays or the
hadronic  component of the photon are expected to contribute only for $z \lsim 0.4$
and are estimated to be negligible.

\begin{figure} \centering
\begin{picture}(16.0,18.3)
  \put(-0.6,-0.5){\epsfig{file=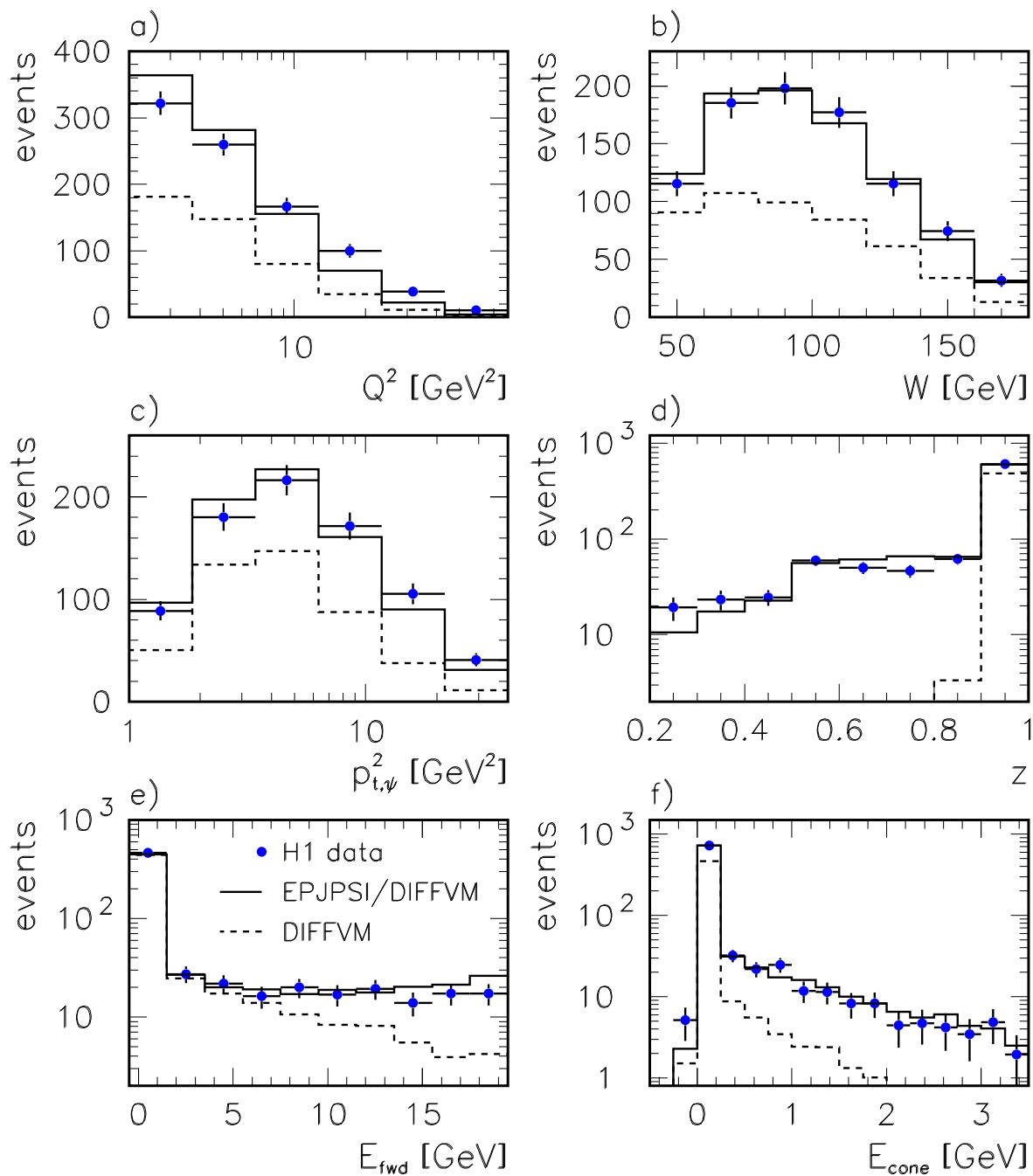,width=18cm}}
\end{picture}
\caption{Comparison between data and Monte Carlo simulations for inclusive
  $J/\psi$ production after all selection  cuts in Table~\protect\ref{cuts} and
  after background subtraction.  Shown are distributions of a) $Q^2$, b) $W$, c)
  the square of the $J/\psi$ transverse momentum in the laboratory frame
  $p_{t,\psi}^2$, d) the elasticity $z$, e) the energy $E_{fwd}$  deposited in the
  LAr calorimeter for $\theta < 20^\circ$, and f) the energy $E_{cone}$  in a cone
  with radius $R = \sqrt{(\Delta\eta)^2+(\Delta\phi)^2}=1$ ($\eta = -\ln\tan
  (\theta / 2)$) around the $J/\psi$ direction of flight. The results of the
  combined Monte Carlo simulation (\mbox{DIFFVM} and EPJPSI, full lines) and of the
  \mbox{DIFFVM} simulation only (dashed lines) are shown. The error bars on the data
  points are statistical only.}
\label{dataMCinela}
\end{figure}

The EPJPSI contribution is normalized to the data in the interval $0.4<z<0.8$
and the \mbox{DIFFVM} contribution is added to describe the data in the region
$z>0.95$ (compare also Fig.~\ref{dataMCinela}d). Numerous checks were carried
out to ensure that all important aspects of the data are well described by
this mix. Comparisons between the data and the Monte Carlo simulation are
shown in  Fig.~\ref{dataMCinela}.

The systematic errors in this analysis are typically $18\%$ and are dominated by
uncertainties in the acceptance corrections, mainly due to the model uncertainty of the
Monte Carlo description, the subtraction of non-resonant background, and
reconstruction efficiencies. The largest systematic uncertainty (up to $32\%$) is
found for large $z$ and small $p_{t,\psi}^2$ values.

\begin{figure} \centering
\begin{picture}(13.4,6.5)
\put(0.0,-0.6){\epsfig{file=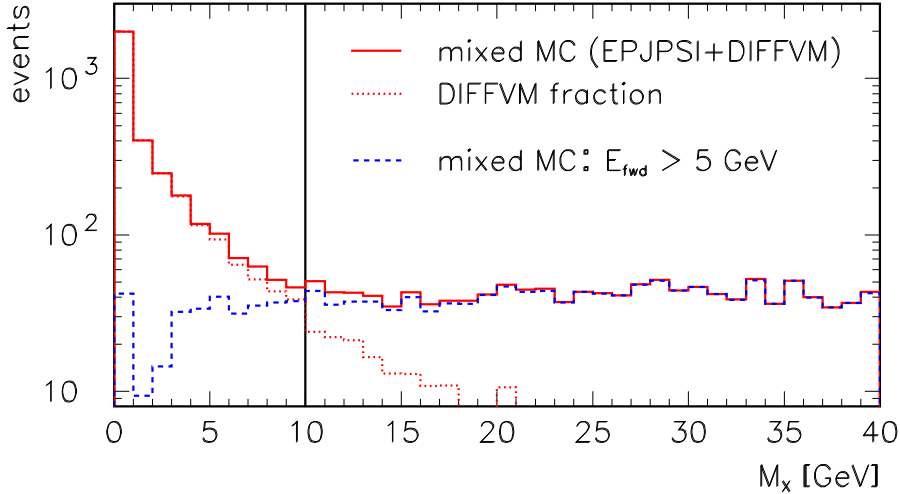,width=13.4cm}}
\end{picture}
\caption{Distribution of simulated $J/\psi$ events as a function of the 
  generated value of the mass 
  $M_X$. The mixed Monte Carlo sample (\mbox{DIFFVM} and \mbox{EPJPSI}) is shown before 
  (full histogram) and after (dashed histogram) applying the cut 
  $E_{fwd}>5\mbox{~GeV}$. The diffractive contribution as simulated by \mbox{DIFFVM}
  before the cut is also shown (dotted histogram).}
\label{mxmc}
\end{figure}

In a second step inelastic cross sections are determined  for events with
a large energy deposition in the forward region of the LAr calorimeter,
namely $E_{fwd}>5\mbox{~GeV}$ for polar angles  $\theta<20^\circ$. This
requirement is an indirect cut on the mass of  the hadronic system $X$.
Its effect can be seen for simulated events in Fig.~\ref{mxmc} where the
$M_X$ distribution is shown for the  mixed simulation (\mbox{EPJPSI+DIFFVM}) and
for the fraction of the diffractive  simulation (\mbox{DIFFVM}) separately. The
latter dominates at low values of $M_X$ and is suppressed efficiently by
the cut on $E_{fwd}$.  The differential cross sections for $M_X >
10\mbox{~GeV}$  are thus determined by applying the cut
$E_{fwd}>5\mbox{~GeV}$ and  then correcting to $M_X > 10\mbox{~GeV}$ using
the Monte Carlo simulation.

\subsection{Differential Cross Sections}

\begin{figure}[p] \centering
\begin{picture}(15.8,18.)
  \put(0.0,-0.9){\epsfig{file=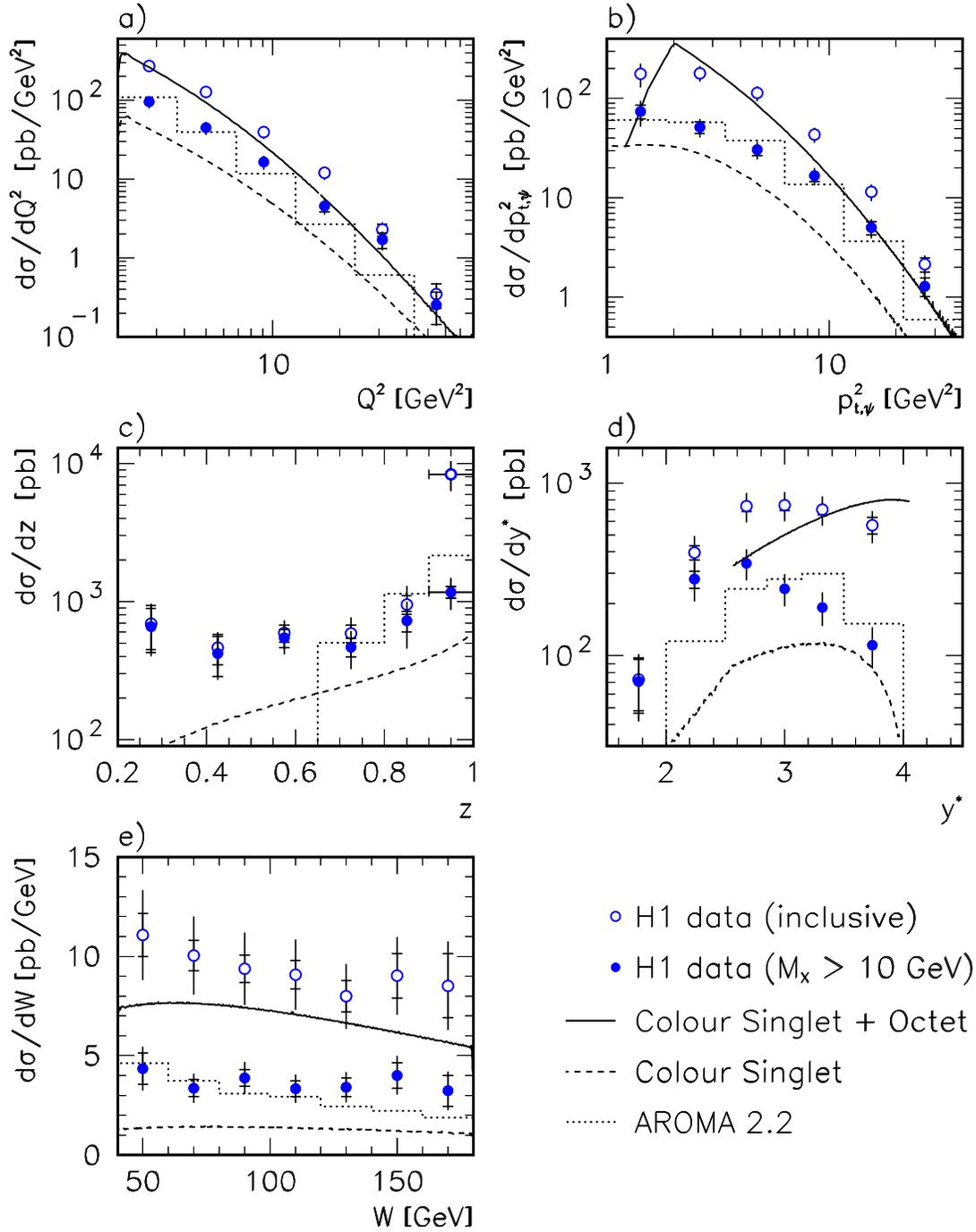,width=16.5cm}}
\end{picture}
\caption{Differential cross sections for the inclusive (open points)
  and inelastic ($M_X>10\mbox{~GeV}$, black points) $ep \rightarrow e\:J/\psi\: X$
  process. a) $d\sigma / dQ^2$, b) $d\sigma / dp_{t,\psi}^2$ (see also footnote
  \ref{edgenote} concerning the theoretical prediction), c) $d\sigma / dz$,  d)
  $d\sigma /dy^\ast$ and e) $d\sigma / dW$.  The kinematic region is $2 < Q^2 <
  80\mbox{~GeV}^2$, $40 < W < 180\mbox{~GeV}$ and $z > 0.2$. The inner error bars are
  statistical, the outer  error bars contain statistical and systematic uncertainties
  added in  quadrature. The dotted histogram gives the prediction from the SCI model
  in AROMA\,2.2 \cite{aroma,rathpriv}  for inclusive $J/\psi$ production. The curves
  are predictions for inelastic $J/\psi$ production within the NRQCD factorization
  approach \protect\cite{Fle972} for the colour singlet contribution (dashed line) and
  the sum of singlet and octet contributions (full line).}
\label{incldiff}
\end{figure}

\paragraph{Inclusive Cross Sections and Soft Colour Interactions}
Differential $ep$ cross sections for inclusive $J/\psi$ production  are given in
Table \ref{summarytable3} and shown in Fig.~\ref{incldiff} (open points) as
functions of $Q^2$, $p_{t,\psi}^2$, $z$, $y^\ast$ (the rapidity of the  $J/\psi$
in the $\gamma^* p$ centre of mass system) and $W$. The prediction of the Soft
Colour Interaction Model (dotted histogram in Fig.~\ref{incldiff}) which is
computed using a modified version of AROMA \cite{aroma,rathpriv}\footnote{The
following parameters are used in addition to standard settings: charm mass
$m_c=1.4\mbox{~GeV}$, GRV(HO) parton densities and $R=0.5$, where $R$
parameterizes the probability of a colour exchange between partons.}, is compared
to the data. Although the model gives a reasonable description of the shapes of
several distributions, there are major discrepancies in the $z$ distribution
and   in the absolute values of the measured and predicted cross sections. For
small $z$, the SCI model as implemented in AROMA is expected to fall below the
data due to the missing hard  contributions of the Colour Singlet Model which
should also be taken into account \cite{rathpriv}. At large $z$ the AROMA SCI
prediction is below the measured cross section by approximately a factor of four.

\paragraph{Inelastic Cross Sections and NRQCD Factorization Approach}

The differential cross sections for inelastic $J/\psi$ production, that is for
$M_X>10\mbox{~GeV}$, are also displayed in Fig.~\ref{incldiff} (full points).
In comparison with the inclusive cross sections  the effect of requiring a high
mass is most significant in the shapes of the distributions of $z$ and
$y^{\ast}$.

The results of the calculations 
by Fleming and Mehen \cite{Fle972} who applied the NRQCD factorization
approach to electroproduction of $J/\psi$ mesons are shown in 
Fig.~\ref{incldiff} for comparison. The predicted cross sections include
the contributions from the colour octet states $^{3}P_0$, $^{1}S_0$ which are
of order ${\cal O}(\alpha_s)$ and the colour singlet state $^{3}S_1$ (of
order ${\cal O}(\alpha_s^2)$)\footnote{
 Spectroscopic notation is used: $^{2S+1}L_J$ where $S$, $L$ and $J$ denote 
 spin, orbital and total angular momentum of the $c\bar{c}$ system that is 
 produced in the hard process. The following values for non-perturbative long
 range transition matrix elements are used:
 $\langle {\cal O}^{J/\psi}_{(1)} (^{3}S_1)\rangle = 1.1\mbox{~GeV}^3$,
 $\langle {\cal O}^{J/\psi}_{(8)} (^{1}S_0)\rangle = 0.01\mbox{~GeV}^3$, and
 $\langle {\cal O}^{J/\psi}_{(8)} (^{3}P_0)\rangle/m_c^2 = 0.005\mbox{~GeV}^3$.
 The octet matrix elements $\langle {\cal O}^{J/\psi}_{(8)}\rangle$ were
 estimated from fits to the CDF data performed in
 \cite{refme} while the singlet matrix element is calculated from the measured
 electronic decay width of the $J/\psi$.}.
The sum of these contributions shown in Fig.~\ref{incldiff} 
is computed using  GRV(LO) \cite{GRV} parton densities; the
colour singlet  contribution is also shown separately. Note that these 
predictions are for $c\bar{c}$ pairs from the hard subprocess and 
do not include any hadronisation effects. The hadronisation of the colour 
octet $c\bar{c}$ pairs into a colour singlet $J/\psi$ is believed to 
proceed via emission of soft gluons \footnote{\label{edgenote}
The sharp edge observed in Fig.~\ref{incldiff}b in the theoretical
$p_{t,\psi}^2$ curve is a consequence of the missing hadronisation.}.

The colour octet contribution dominates the cross section for all $Q^2$
(Fig.~\ref{incldiff}a). The colour singlet contribution (dashed curves in
Fig.~\ref{incldiff}) is seen to fall below the data by factors $2 - 3$ while the
prediction for the sum (full curves in Fig.~\ref{incldiff}) is overall too large in
absolute magnitude by up to a factor 3. The shapes of the data distributions are
not well reproduced by the calculation: the predicted $Q^2$ and $p_{t,\psi}^2$
distributions are steeper than the data and the $y^\ast$ distribution increases
towards larger values of $y^\ast$ instead of falling. The $W$ distribution  agrees
in shape but overshoots the data. The full prediction of the $z$ distribution is at
present not calculable \cite{Fle972} and is therefore not shown.

The observed differences in magnitude between the predicted and measured  cross
sections probably call for an overall adjustment of the fitted transition
matrix elements while the shapes may be influenced by a relative adjustment of
the individual contributions. There is a hint that these differences  increase
towards low $Q^2$ (see Fig.~\ref{incldiff}a, full points and full curve). The
theoretical predictions are also expected to be more precise for larger $Q^2$
and $p_{t,\psi}^2$ \cite{Fle972}. Therefore the comparison was repeated for
$Q^2 > 4\mbox{~GeV}^2$ and $p_{t,\psi}^2 > 4\mbox{~GeV}^2$ (see Table
\ref{summarytable4}), but no significant change in the conclusions was found.

\subsection{Integrated Cross Sections and Comparison with Photoproduction}
 
In Table \ref{ics} the integrated cross sections for  $e+p \rightarrow e+
J/\psi+ X$ in the kinematic region $2 < Q^2 < 80\mbox{~GeV}^2$, $40 < W <
180\mbox{~GeV}$ and $z > 0.2$ are summarised. They are given for the inclusive
selection,  for the inelastic selection corresponding to  $M_X>10\mbox{~GeV}$,
and for $z < 0.9$ as in previous photoproduction  analyses
\cite{H1962,Zeus972}. In addition, the cross sections after imposing the
additional cuts $Q^2 > 4.0\mbox{~GeV}^2$ and $p_{t,\psi}^2 > 4.0\mbox{~GeV}^2$
are provided.

\begin{table}[!t] \centering
\begin{tabular}{|l|c|c|} \hline
      &   \multicolumn{2}{|c|}{$\sigma (ep \rightarrow e\: J/\psi\: X)$ \ \ \ [nb]} \\[.1em]
    \rb{Data set} & \ \ \ \ \ \ $Q^2 > 2.0\mbox{~GeV}^2$ \ \ \ \ \ \ &
                        $Q^2$,\,$p_{t,\psi}^2 > 4.0\mbox{~GeV}^2$\\[.2em] \hline
    Inclusive             & $1.30\pm0.06\pm0.24$ &  $0.50\pm0.04\pm0.09$ \\
    $M_X>10\mbox{~GeV}$   & $0.51\pm0.04\pm0.09$ &  $0.17\pm0.02\pm0.03$ \\
    $z<0.9$               & $0.46\pm0.04\pm0.09$ &  $0.13\pm0.02\pm0.02$ \\ \hline
\end{tabular}
\caption{Integrated cross sections for $40 < W < 180\mbox{~GeV}$, 
  $Q^2 < 80\mbox{~GeV}^2$ and $z > 0.2$ in two kinematic regions,  
  $Q^2 > 2.0\mbox{~GeV}^2$ and $Q^2$ and $p_{t,\psi}^2 > 4.0\mbox{~GeV}^2$.
  Results are given for the inclusive 
  selection and for two inelastic selections, $M_X>10\mbox{~GeV}$ and $z<0.9$.}
  \label{ics}
\end{table}

\begin{figure}[!p] \centering
\begin{picture}(10.8,8.5)
  \put(0.0,-0.7){\epsfig{file=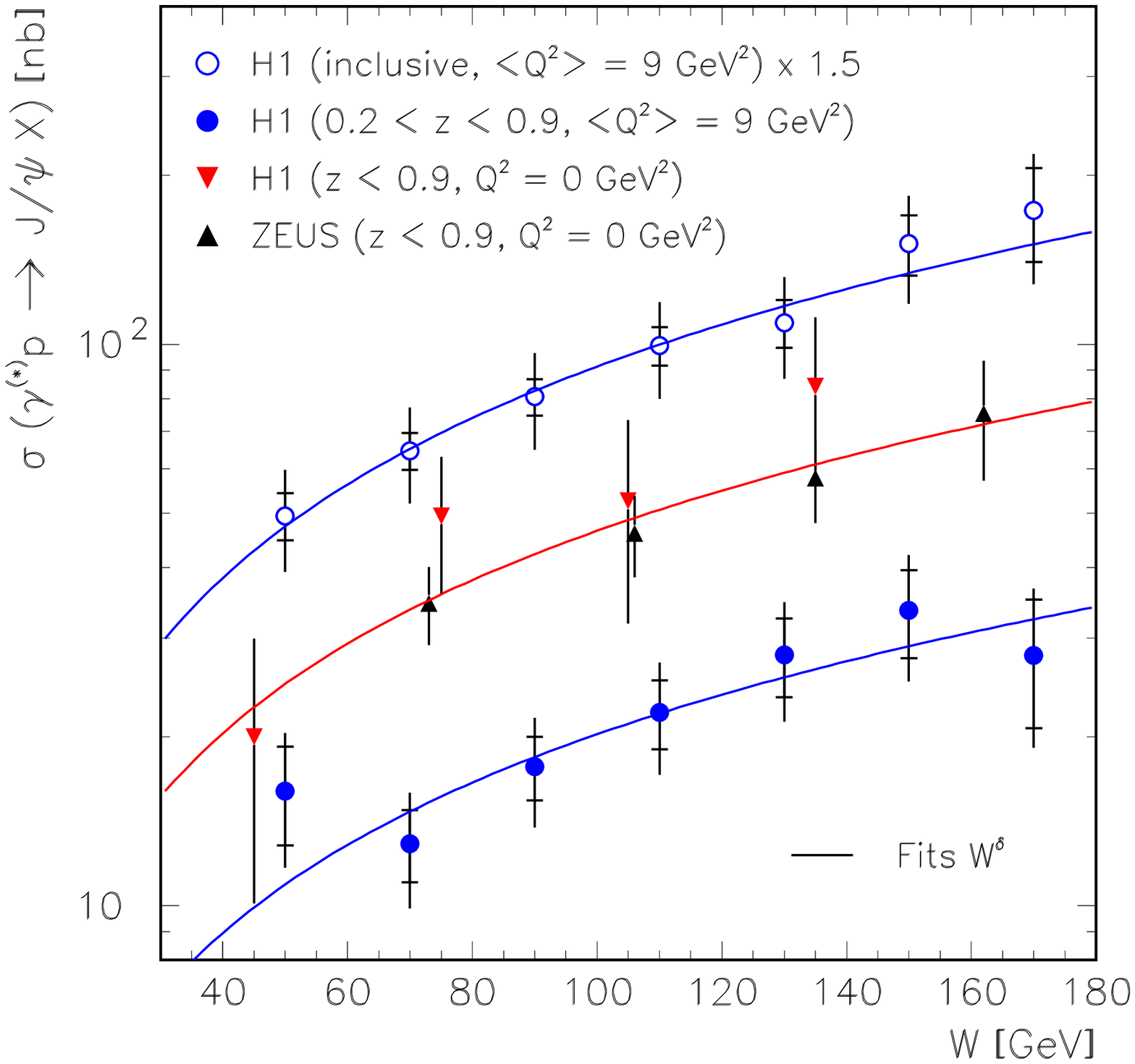,width=10.8cm}}
\end{picture}
\caption{Total cross sections for $\gamma^{\ast}\,p\rightarrow J/\psi \,X$
  from this analysis at $\langle Q^2\rangle = 9\mbox{~GeV}^2$.  The inclusive cross section
  ($0.2 < z \lsim 1.0$) is shown as a function  of $W$ (multiplied by a factor 1.5 for
  clarity), as well as the cross section for $0.2 < z < 0.9$. Photoproduction data
  \cite{H1962,Zeus972} with similar cuts in $z$ are included for comparison. The inner
  error bars on the points from this analysis indicate the statistical uncertainty, while
  the outer bars show the statistical and systematic uncertainties added in quadrature.}
\label{inclw}
\begin{picture}(10.2,8.1)
  \put(0.0,-1.0){\epsfig{file=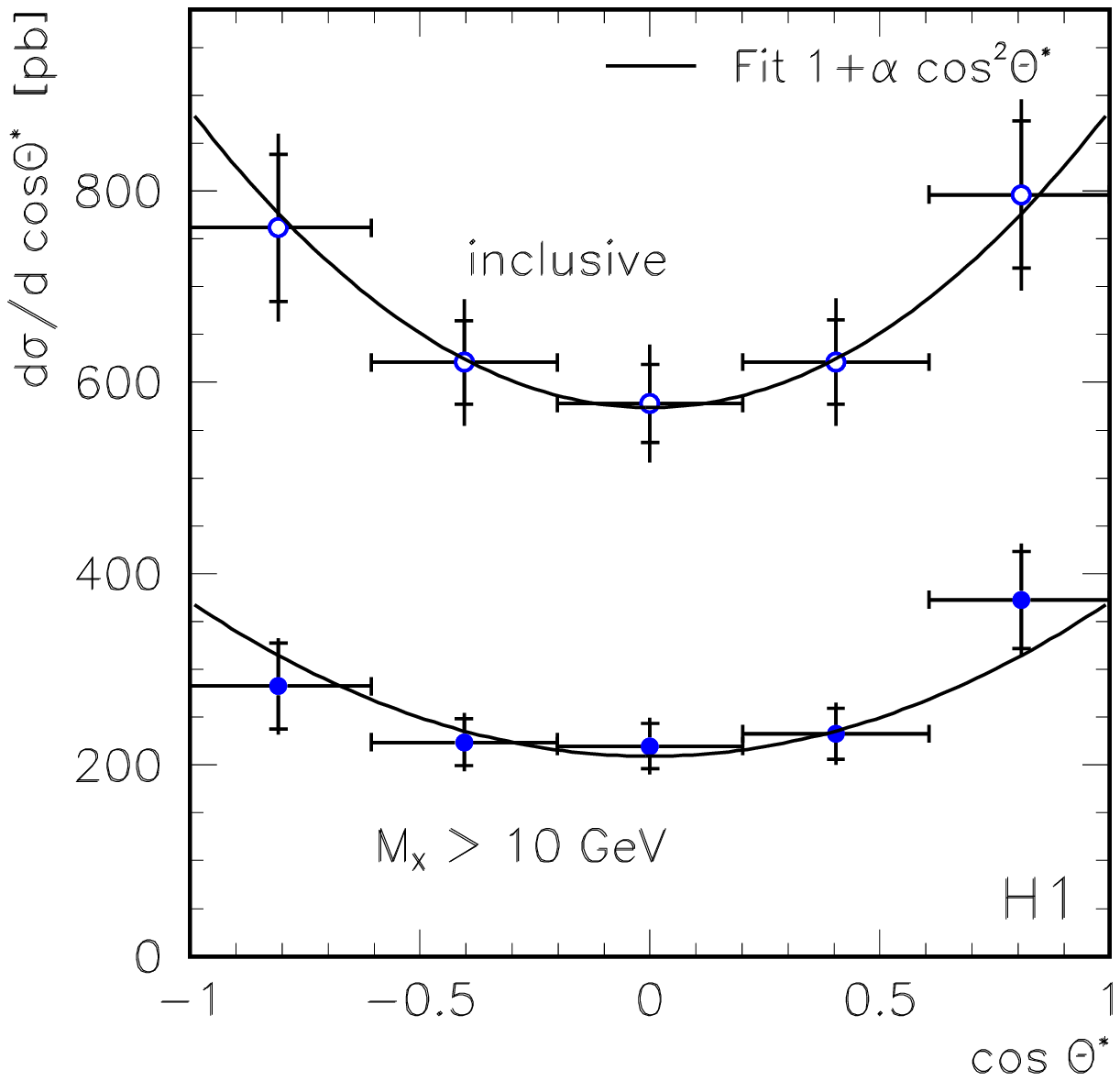,width=10.2cm}}
\end{picture}
\caption{Differential cross sections $d\sigma / d\cos\theta^\ast$ for 
  $ep \rightarrow e\: J/\psi\: X$ in the kinematic region $2 < Q^2 <
  80\mbox{~GeV}^2$,  $40 < W < 180\mbox{~GeV}$ and $z > 0.2$. The inclusive
  cross section and the inelastic cross section   ($M_X>10\mbox{~GeV}$) are
  shown. The inner error bars indicate the statistical  uncertainty, while
  the outer bars include the statistical and  systematic uncertainties added
  in quadrature. The lines are fits to the form $\sim 1+\alpha
  \cos^2\theta^\ast$.}
 \label{thetastar}
\end{figure}

The total cross section for $\gamma^* p \rightarrow J/\psi\,X$ is computed
according to equation (\ref{eq:sigma}) as a function of the $\gamma^* p$ centre
of mass energy $W$ at $\langle Q^2\rangle = 9\mbox{~GeV}^2$ and is given in
Fig.~\ref{inclw} and Table \ref{summarytable5}. The cross section is determined
for the  inclusive data $0.2 < z \lsim 1$ and, in view of a comparison with
photoproduction, also for an inelastic selection using a  cut $z<0.9$
\footnote{The photoproduction data are given for $0<z<0.9$. This was achieved by
an extrapolation to $z=0$  assuming contributions from photon gluon fusion
only.  This contribution at small $z$ is however negligible in the comparison.}.
The $W$ dependence is seen to be very similar to that in  the photoproduction
data \cite{H1962,Zeus972} also shown in Fig.~\ref{inclw}. The $W$ dependence,
parameterized as $W^{\delta}$, yields $\delta = 0.95 \pm 0.11$  for the
inclusive data and $\delta = 0.89 \pm 0.20$ for the data with $z < 0.9$. The
photoproduction data, including H1 and ZEUS, are described by $\delta = 0.91 \pm
0.26$. The errors include statistical and systematic uncertainties.

\subsection{Decay Angular Distributions}

Measuring the polarization of the $J/\psi$ is thought to be a way of
distinguishing the various contributions to $J/\psi$ production. The  polar
($\theta^\ast$) decay angular distributions in the helicity frame are shown in
Fig.~\ref{thetastar} for the fully inclusive case and for the inelastic
selection $M_X>10\mbox{~GeV}$, in the kinematic region $2 < Q^2 <
80\mbox{~GeV}^2$, $40 < W < 180\mbox{~GeV}$ and $z > 0.2$.

The $\cos\theta^\ast$ distribution is predicted to have the form
\begin{equation}
  \frac{{\rm d}\,\sigma}{{\rm d}\,\cos\theta^{\ast}}\propto 1 + \alpha\, \cos^2\theta^{\ast}.
\end{equation}

For $J/\psi$ production via the colour singlet mechanism   $\alpha\approx 0.5$
is expected for the kinematic range studied here \cite{flepriv}. If colour
octet contributions are present, $|\alpha|\lsim 0.5$ is expected, where
$\alpha$  can be negative, zero or positive depending on which intermediate
$c\bar{c}$ state dominates the production \cite{Fle972}.

The data yield values of $\alpha = 0.54^{+0.29}_{-0.26}$ for the inclusive case and
$\alpha = 0.77^{+0.44}_{-0.38}$ for the inelastic selection ($M_X >
10\mbox{~GeV}$), including statistical and systematic uncertainties. The
uncertainties are too large to draw definite conclusions.


\section{Summary and Conclusions}

Measurements of elastic $J/\psi$ production in deep inelastic scattering (DIS) with $25 <
W < 160\mbox{~GeV}$ and $2 < Q^2 < 80\mbox{~GeV}^2$ have been presented. They are more
precise and cover a larger kinematic range than previous analyses at HERA. The dependence
of the cross section $\sigma(\gamma^* p \rightarrow J/\psi\: p)$ on $W$ is found to be
proportional to $W^\delta$ with $\delta \simeq 1$, as was also observed in photoproduction.
The $Q^2$ dependence is measured to be $\propto 1/(Q^2+m_\psi^2)^n$ with $n = 2.38 \pm 0.11$. 
Both the $W$ and $Q^2$ dependence are well described by a model based on perturbative QCD.

Assuming that the $t$ dependence of elastic $J/\psi$ production can be described by
one exponential distribution, the slope parameter is determined to be
$b = 4.1 \pm 0.3\mbox{~(stat.)}\pm 0.4\mbox{~(syst.)}\mbox{~GeV}^{-2}$, compatible
with the value found in photoproduction. The helicity structure of quasi-elastic $J/\psi$
production in DIS has been investigated and no evidence is found for a violation of
$s$-channel helicity conservation. Assuming SCHC the ratio $R$ of the longitudinal to
the transverse cross section has been determined using the $\cos\theta^\ast$
distribution in two $Q^2$ regimes; the result is $R = 0.18^{+0.18}_{-0.14}$
for $\langle Q^2\rangle = 4\mbox{~GeV}^2$ and $R = 0.94^{+0.79}_{-0.43}$ for $\langle
Q^2\rangle = 16\mbox{~GeV}^2$, suggesting a rise with $Q^2$.

The first evidence from HERA for quasi-elastic production of $\psi(2S)$ mesons in DIS
has been reported. The increase of the ratio of cross sections for $\psi(2S)$ and
$J/\psi$ production with $Q^2$ predicted by models is supported by the data.

Data have been presented for the inclusive production of $J/\psi$ mesons in deep
inelastic scattering, covering the kinematic region $40 < W < 180\mbox{~GeV}$, $2 <
Q^2 < 80\mbox{~GeV}^2$ and $0.2 < z \lsim 1$. Differential $ep$ cross
sections are computed as functions of $Q^2$, $p_{t,\psi}^2$, $z$,
$y^\ast$ and $W$. The model of Soft Colour Interactions, a non-perturbative
phenomenological approach to the description of inclusive $J/\psi$ production,
is compared to the data. The dependences of the differential cross sections on several
variables are reasonably well described by the model, but the normalisations and
the $z$ dependence are not reproduced.

Using a selection cut designed to reject events with a low mass hadronic system,
diffractive events are suppressed  and inelastic cross sections ($M_X >
10\mbox{~GeV}$) are extracted. A leading order calculation in the NRQCD
factorization approach using long range matrix elements determined from $J/\psi$
production in $p\bar{p}$ collisions at the Tevatron is confronted with our
measurements of differential inelastic cross sections. The shape and magnitude of
the differential distributions are not described by the theoretical prediction.
The comparisons may indicate the need to decrease the size of the colour octet
long distance matrix elements or to change the relative importance of the
different colour octet contributions, and/or to include  higher orders in the
NRQCD perturbative expansion. The colour singlet contribution alone is below the
data by factors $2 - 3$.


\section*{Acknowledgements}
We are grateful to the HERA machine group whose outstanding efforts made this experiment
possible. We appreciate the immense effort of the engineers and technicians who
constructed and maintained the detector. We thank the funding agencies for their
financial support of the experiment. We wish to thank the DESY directorate for the
support and hospitality extended to the non-DESY members of the collaboration. We thank
S.\,Fleming, L.\,Frankfurt, W.\,Koepf, T.\,Mehen, J.\,Rathsman and M.\,Strikman for
valuable discussions and for making their theoretical predictions available.

This paper was submitted for publication very
soon after the tragic and untimely death of the DESY Director Bj\"orn Wiik.
We would like to record here our appreciation of the friendship,
encouragement, and support which he gave at all times to the H1
experiment.


\begin{table}[p] \centering
\begin{small}
\begin{tabular}{c|rcr|r||rll} \hline
  $\langle Q^2\rangle $ & \multicolumn{3}{|c|}{$W$ interval} & $\langle W\rangle$\ \ &
      \multicolumn{3}{c}{$\sigma (\gamma^\ast p \rightarrow J/\psi p)$} \\
  $[\mbox{GeV}^2]$ & \multicolumn{3}{|c|}{$[\mbox{GeV}]$} & $[\mbox{GeV}]$ &
      \multicolumn{3}{c}{$[\mbox{nb}]$} \\\hline
                  & 25 & -- & 40\ \   & 32.0  \  & \ \ 11.7  & $\pm$ 3.2 (stat.) & $\pm$ 2.9 (syst.) \\
                  & 40 & -- & 60\ \   & 49.3  \  &      22.5  & $\pm$ 4.0        & $\pm$ 3.6  \\
                  & 60 & -- & 80\ \   & 69.5  \  &      26.3  & $\pm$ 3.9        & $\pm$ 4.2  \\
  \rb{3.5}       & 80 & -- & 100\ \   & 89.6  \  &     33.1  & $\pm$ 5.1        & $\pm$ 5.2  \\
                  & 100 & -- & 120\ \ & 109.6  \ &      30.7  & $\pm$ 5.8        & $\pm$ 4.8  \\
                  & 120 & -- & 160\ \ & 138.6  \ &      54.9  & $\pm$ 8.9        & $\pm$ 8.8 \\\hline
                  & 40 & -- & 80\ \   & 57.5   \ &      5.4   & $\pm$ 1.3        & $\pm$ 0.9  \\
  10.1            & 80 & -- & 120\ \  & 98.4   \ &      10.3  & $\pm$ 2.2        & $\pm$ 1.7  \\
                  & 120 & -- & 160\ \ & 138.6  \ &      17.8  & $\pm$ 4.2        & $\pm$ 3.1  \\\hline
  33.6            & 40 & -- & 160\ \  &84.4   \ &      1.34  & $\pm$ 0.37       & $\pm$ 0.24 \\\hline
\end{tabular}
\caption{Cross sections for the elastic process $\gamma^\ast p \rightarrow J/\psi\: p$ in bins of
  $W$ for three $Q^2$ regions: $2 < Q^2 < 6\mbox{~GeV}^2$, $6 < Q^2 < 18\mbox{~GeV}^2$ and
  $18 < Q^2 < 80\mbox{~GeV}^2$.}
\label{summarytable1}
\end{small}
\end{table}

\begin{table}[p] \centering
\begin{small}
\begin{tabular}{rcr|r||rll} \hline
  \multicolumn{3}{c|}{$Q^2$ interval} & $\langle Q^2\rangle$\ \ \  &
     \multicolumn{3}{c}{$\sigma (\gamma^\ast p \rightarrow J/\psi p)$\ \ ($W=90\mbox{~GeV}$)} \\
  \multicolumn{3}{c|}{$[\mbox{GeV}^2]$} & $[\mbox{GeV}^2]$ & \multicolumn{3}{c}{$[\mbox{nb}]$} \\\hline
  $2.0$  & -- & $3.2$\ \   &  2.6\ \     & \ \ 31.9   & $\pm$ 2.5 (stat.) & $\pm$ 5.1 (syst.) \\
  $3.2$  & -- & $5.0$\ \   &  4.1\ \     &      26.8   & $\pm$ 2.4         & $\pm$  4.2 \\
  $5.0$  & -- & $8.0$\ \   &  6.4\ \     &      17.2   & $\pm$ 1.7         & $\pm$  3.0 \\
  $8.0$  & -- & $12.7$\ \  & 10.1\ \     &      11.5   & $\pm$ 1.3         & $\pm$  2.0 \\
  $12.7$ & -- & $20.1$\ \  & 16.0\ \     &       6.4   & $\pm$ 1.0         & $\pm$  1.1 \\
  $20.1$ & -- & $31.8$\ \  & 25.0\ \     &       2.20  & $\pm$ 0.55        & $\pm$ 0.38 \\
  $31.8$ & -- & $80.0$\ \  & 50.0\ \     &       0.57  & $\pm$ 0.20        & $\pm$ 0.10 \\\hline
\end{tabular}
\caption{$Q^2$ dependence of the elastic cross section $\sigma(\gamma^\ast p \rightarrow J/\psi\: p)$.}
\label{summarytable2}
\end{small}
\end{table}

\begin{table}[p] \centering
\begin{small}
\begin{tabular}{r||rll|rll} \hline
  $Q^2$\ \ \ & \multicolumn{6}{c}{\mbox{d}\,$\sigma (e p \rightarrow e J/\psi X) / \mbox{d}\,Q^2$
                     \ \ \ \ \ $[\mbox{pb}/\mbox{GeV}^2]$} \\
  $[\mbox{GeV}^2]$ & \multicolumn{3}{c|}{inclusive} & \multicolumn{3}{c}{$M_X > 10\mbox{~GeV}$} \\\hline
   2.8\ \     & \ \ 269  & $\pm$ 16  (stat.) & $\pm$ 43  (syst.) & \ \ 95.5 & $\pm$ 8.8  (stat.) & $\pm$ 15.3  (syst.) \\
   5.0\ \     & 127       & $\pm$ 8          & $\pm$ 20          & 44.4      & $\pm$ 4.3          & $\pm$ 7.1  \\
   9.1\ \     & 39.5      & $\pm$ 3.2        & $\pm$ 6.3         & 16.4      & $\pm$ 2.0          & $\pm$ 2.6  \\
  17.1\ \     & 12.0      & $\pm$ 1.3        & $\pm$ 1.9         & 4.5       & $\pm$ 0.7          & $\pm$ 0.7  \\
  31.1\ \     & 2.28      & $\pm$ 0.40       & $\pm$ 0.36        & 1.69      & $\pm$ 0.37         & $\pm$ 0.27 \\
  54.5\ \     & 0.35      & $\pm$ 0.12       & $\pm$ 0.06        & 0.25      & $\pm$ 0.11         & $\pm$ 0.04 \\\hline
  $p_{t,\psi}^2$\ \ \ & \multicolumn{6}{c}{\mbox{d}\,$\sigma (e p \rightarrow e J/\psi X) / \mbox{d}\,p_{t,\psi}^2$
                     \ \ \ \ \ $[\mbox{pb}/\mbox{GeV}^2]$} \\
  $[\mbox{GeV}^2]$ & \multicolumn{3}{c|}{inclusive} & \multicolumn{3}{c}{$M_X > 10\mbox{~GeV}$} \\\hline
   1.4\ \     & \ \ 177  & $\pm$ 20  (stat.) & $\pm$ 44  (syst.) & \ \ 73.7 & $\pm$ 11.6  (stat.) & $\pm$ 18.4  (syst.) \\
   2.6\ \     & 180       & $\pm$ 14         & $\pm$ 29          & 51.4      & $\pm$ 7.0          & $\pm$ 8.2 \\
   4.8\ \     & 114       & $\pm$ 8          & $\pm$ 18          & 30.5      & $\pm$ 3.8          & $\pm$ 4.9 \\
   8.6\ \     & 43.4      & $\pm$ 3.5        & $\pm$ 6.9         & 16.7      & $\pm$ 2.1          & $\pm$ 2.7 \\
  15.5\ \     & 11.5      & $\pm$ 1.2        & $\pm$ 1.8         & 5.0       & $\pm$ 0.7          & $\pm$ 0.8 \\
  27.0\ \     & 2.14      & $\pm$ 0.35       & $\pm$ 0.34        & 1.28      & $\pm$ 0.27         & $\pm$ 0.20 \\\hline
     & \multicolumn{6}{c}{\mbox{d}\,$\sigma (e p \rightarrow e J/\psi X) / \mbox{d}\,z$
                     \ \ \ \ \ $[\mbox{pb}]$} \\
  \rb{$z$\ \ \ \ } & \multicolumn{3}{c|}{inclusive} & \multicolumn{3}{c}{$M_X > 10\mbox{~GeV}$} \\\hline
   0.275\ \     & \ \ 690  & $\pm$ 240  (stat.) & $\pm$ 110  (syst.) & \ \ 660 & $\pm$ 230  (stat.) & $\pm$ 110  (syst.) \\
   0.425\ \     & 460       & $\pm$ 110         & $\pm$ 70          & 420      & $\pm$ 130          & $\pm$ 70  \\
   0.575\ \     & 590       & $\pm$ 80          & $\pm$ 110         & 550      & $\pm$ 80           & $\pm$ 100 \\
   0.725\ \     & 590       & $\pm$ 80          & $\pm$ 160         & 470      & $\pm$ 70           & $\pm$ 130 \\
   0.850\ \     & 950       & $\pm$ 150         & $\pm$ 310         & 730      & $\pm$ 130          & $\pm$ 240 \\
   0.950\ \     & 8350      & $\pm$ 390         & $\pm$ 2000        & 1170     & $\pm$ 110          & $\pm$ 280 \\\hline
      & \multicolumn{6}{c}{\mbox{d}\,$\sigma (e p \rightarrow e J/\psi X) / \mbox{d}\,y^\ast$
                     \ \ \ \ \ $[\mbox{pb}]$} \\
  \rb{$y^\ast$\ \ \ } & \multicolumn{3}{c|}{inclusive} & \multicolumn{3}{c}{$M_X > 10\mbox{~GeV}$} \\\hline
   1.77\ \     & \ \ 73   & $\pm$ 25  (stat.) & $\pm$ 17  (syst.) & \ \ 71 & $\pm$ 24  (stat.) & $\pm$ 16  (syst.) \\
   2.24\ \     & 396       & $\pm$ 38          & $\pm$ 91         & 277      & $\pm$ 31          & $\pm$ 64  \\
   2.68\ \     & 737       & $\pm$ 53          & $\pm$ 133        & 343      & $\pm$ 35          & $\pm$ 62 \\
   3.00\ \     & 746       & $\pm$ 51          & $\pm$ 134        & 244      & $\pm$ 28          & $\pm$ 44 \\
   3.32\ \     & 702       & $\pm$ 49          & $\pm$ 126        & 191      & $\pm$ 25          & $\pm$ 34 \\
   3.74\ \     & 568       & $\pm$ 62          & $\pm$ 102        & 115      & $\pm$ 23          & $\pm$ 21 \\\hline
  $W$\ \ \ \ & \multicolumn{6}{c}{\mbox{d}\,$\sigma (e p \rightarrow e J/\psi X) / \mbox{d}\,W$
                     \ \ \ \ \ $[\mbox{pb}/\mbox{GeV}]$} \\
  $[\mbox{GeV}]$  & \multicolumn{3}{c|}{inclusive} & \multicolumn{3}{c}{$M_X > 10\mbox{~GeV}$} \\\hline
   50\ \     & \ \ 11.1  & $\pm$ 1.1  (stat.) & $\pm$ 2.0  (syst.) & \ \ 4.3 & $\pm$ 0.8  (stat.) & $\pm$ 0.8  (syst.) \\
   70\ \     & 10.0      & $\pm$ 0.8          & $\pm$ 1.8         & 3.4      & $\pm$ 0.4          & $\pm$ 0.6 \\
   90\ \     & 9.4       & $\pm$ 0.7          & $\pm$ 1.7        & 3.9      & $\pm$ 0.4          & $\pm$ 0.7 \\
   110\ \    & 9.1       & $\pm$ 0.7          & $\pm$ 1.6        & 3.3      & $\pm$ 0.4          & $\pm$ 0.6 \\
   130\ \    & 8.0       & $\pm$ 0.8          & $\pm$ 1.4        & 3.4      & $\pm$ 0.5          & $\pm$ 0.6 \\
   150\ \    & 9.0       & $\pm$ 1.1          & $\pm$ 1.6        & 4.0      & $\pm$ 0.6          & $\pm$ 0.7 \\
   170\ \    & 8.5       & $\pm$ 1.6          & $\pm$ 1.5        & 3.2      & $\pm$ 0.8          & $\pm$ 0.6 \\\hline
\end{tabular}
\caption{Inclusive and inelastic ($M_X>10\,$GeV) differential cross sections for the 
  process $e p \rightarrow e\: J/\psi\: X$ in the kinematic region
  $2<Q^2<80\mbox{~GeV}^2$, $40 < W < 180\mbox{~GeV}$ and $z>0.2$.}
\label{summarytable3}
\end{small}
\end{table}

\begin{table}[p] \centering
\begin{small}
\begin{tabular}{r||rll|rll} \hline
     & \multicolumn{6}{c}{\mbox{d}\,$\sigma (e p \rightarrow e J/\psi X) / \mbox{d}\,z$
                     \ \ \ \ \ $[\mbox{pb}]$} \\
  \rb{$z$\ \ \ \ } & \multicolumn{3}{c|}{inclusive} & \multicolumn{3}{c}{$M_X > 10\mbox{~GeV}$} \\\hline
   0.275\ \     & \ \ 110  & $\pm$  80  (stat.) & $\pm$ 20  (syst.) & \ \ 110 & $\pm$ 80  (stat.) & $\pm$ 20  (syst.) \\
   0.425\ \     & 210       & $\pm$ 80         & $\pm$ 30          & 190       & $\pm$ 70           & $\pm$ 30 \\
   0.575\ \     & 180       & $\pm$ 50         & $\pm$ 30          & 160       & $\pm$ 50           & $\pm$ 30 \\
   0.725\ \     & 220       & $\pm$ 50         & $\pm$ 60          & 200       & $\pm$ 50           & $\pm$ 50 \\
   0.850\ \     & 250       & $\pm$ 70         & $\pm$ 80          & 190       & $\pm$ 70           & $\pm$ 60 \\
   0.950\ \     & 3750      & $\pm$ 240        & $\pm$ 900         & 530       & $\pm$ 70           & $\pm$ 130 \\\hline
      & \multicolumn{6}{c}{\mbox{d}\,$\sigma (e p \rightarrow e J/\psi X) / \mbox{d}\,y^\ast$
                     \ \ \ \ \ $[\mbox{pb}]$} \\
  \rb{$y^\ast$\ \ \ } & \multicolumn{3}{c|}{inclusive} & \multicolumn{3}{c}{$M_X > 10\mbox{~GeV}$} \\\hline
   1.77\ \     & \ \ 15   & $\pm$ 10  (stat.) & $\pm$ 3  (syst.) & \ \ 15 & $\pm$ 10  (stat.) & $\pm$ 3  (syst.) \\
   2.24\ \     & 155       & $\pm$ 22          & $\pm$ 36        & 95      & $\pm$ 17          & $\pm$ 22 \\
   2.68\ \     & 328       & $\pm$ 33          & $\pm$ 59        & 126     & $\pm$ 20          & $\pm$ 23 \\
   3.00\ \     & 317       & $\pm$ 32          & $\pm$ 57        & 80      & $\pm$ 15          & $\pm$ 14 \\
   3.32\ \     & 309       & $\pm$ 34          & $\pm$ 56        & 66      & $\pm$ 14          & $\pm$ 12 \\
   3.74\ \     & 169       & $\pm$ 35          & $\pm$ 30        & 31      & $\pm$ 13          & $\pm$ 6 \\\hline
  $W$\ \ \ \ & \multicolumn{6}{c}{\mbox{d}\,$\sigma (e p \rightarrow e J/\psi X) / \mbox{d}\,W$
                     \ \ \ \ \ $[\mbox{pb}/\mbox{GeV}]$} \\
  $[\mbox{GeV}^2]$  & \multicolumn{3}{c|}{inclusive} & \multicolumn{3}{c}{$M_X > 10\mbox{~GeV}$} \\\hline
   50\ \     & \ \ 3.8  & $\pm$ 0.6  (stat.) & $\pm$ 0.7  (syst.) & \ \ 1.32 & $\pm$ 0.42  (stat.) & $\pm$ 0.24  (syst.) \\
   70\ \     & 4.2       & $\pm$ 0.5          & $\pm$ 0.8         & 1.12      & $\pm$ 0.25          & $\pm$ 0.20 \\
   90\ \     & 3.9       & $\pm$ 0.4          & $\pm$ 0.7         & 1.32      & $\pm$ 0.23          & $\pm$ 0.24 \\
   110\ \    & 4.1       & $\pm$ 0.5          & $\pm$ 0.7         & 1.30      & $\pm$ 0.24          & $\pm$ 0.23 \\
   130\ \    & 3.2       & $\pm$ 0.5          & $\pm$ 0.6         & 1.12      & $\pm$ 0.25          & $\pm$ 0.20 \\
   150\ \    & 3.5       & $\pm$ 0.6          & $\pm$ 0.6         & 1.63      & $\pm$ 0.35          & $\pm$ 0.29 \\
   170\ \    & 2.1       & $\pm$ 0.6          & $\pm$ 0.4         & 0.80      & $\pm$ 0.33          & $\pm$ 0.14 \\\hline
\end{tabular}
\caption{Inclusive and inelastic ($M_X>10\,$GeV) differential cross sections for the
  process $e p \rightarrow e\: J/\psi\: X$ in the kinematic region
  $4<Q^2<80\mbox{~GeV}^2$, $p_{t,\psi}^2 > 4\mbox{~GeV}^2$, $40 < W < 180\mbox{~GeV}$ and $z>0.2$.}
\label{summarytable4}
\end{small}
\end{table}

\begin{table}[p] \centering
\begin{small}
\begin{tabular}{r||rll|rll} \hline
  $W$\ \ \  & \multicolumn{6}{c}{$\sigma (\gamma^\ast p \rightarrow J/\psi X)$
              \ \ \ \ $[\mbox{nb}]$} \\
  $[\mbox{GeV}]$ & \multicolumn{3}{c|}{inclusive} &  \multicolumn{3}{c}{$z < 0.9$}\\\hline
   50\ \    & \ \ 33.0   & $\pm$ 3.2 (stat.) & $\pm$ 5.9 (syst.) & \ \ 16.0   & $\pm$ 3.2 (stat.) & $\pm$ 2.9 (syst.)\\
   70\ \    &      43.1  & $\pm$ 3.3         & $\pm$  7.8        & \ \ 12.9   & $\pm$ 1.9          & $\pm$ 2.3 \\
   90\ \    &      53.8  & $\pm$ 4.0         & $\pm$  9.7        & \ \ 17.7   & $\pm$ 2.3          & $\pm$ 3.2 \\
  110\ \    &      66.3  & $\pm$ 5.2         & $\pm$ 11.9        & \ \ 22.1   & $\pm$ 3.1          & $\pm$ 4.0 \\
  130\ \    &      72.8  & $\pm$ 7.1         & $\pm$ 13.1        & \ \ 28.0   & $\pm$ 4.5          & $\pm$ 5.0 \\
  150\ \    &     101    & $\pm$ 12          & $\pm$ 18          & \ \ 33.6   & $\pm$ 6.0          & $\pm$ 6.0 \\
  170\ \    &     115    & $\pm$ 22          & $\pm$ 21          & \ \ 27.9   & $\pm$ 7.2          & $\pm$ 5.0 \\\hline
\end{tabular}
\caption{$W$ dependence of the inclusive and the inelastic 
($z<0.9$) cross sections $\sigma(\gamma^\ast p \rightarrow J/\psi\: X)$ in the kinematic region
  $2<Q^2<80\mbox{~GeV}^2$, $40 < W < 180\mbox{~GeV}$ and $z>0.2$.}
\label{summarytable5}
\end{small}
\end{table}

\end{document}